%% file: Bulge_GCs.tex
\documentclass[fleqn,usenatbib,useAMS]{mnras}

\usepackage{newtxtext,newtxmath}

\usepackage[T1]{fontenc}
\usepackage{ae,aecompl}
 
\usepackage{graphicx}	
\usepackage{amsmath}	
\usepackage{amssymb}	
\usepackage{longtable}
\usepackage{natbib}
\usepackage{multirow}
\usepackage{lscape} 
\usepackage{tabularx}

\title[Globular cluster classification]{Globular clusters in the inner Galaxy classified from dynamical orbital criteria}

\author[P\'erez-Villegas et al.]{
Angeles  P\'erez-Villegas,$^{1}$\thanks{E-mail: mperez@iag.usp.br}
Beatriz Barbuy,$^{1}$
Leandro Kerber,$^{2}$
Sergio Ortolani$^{3}$
\newauthor
Stefano O. Souza $^{1}$
and Eduardo Bica,$^{4}$  
\\
$^{1}$Universidade de S\~ao Paulo, IAG, Rua do Mat\~ao 1226, Cidade Universit\'aria, S\~ao Paulo 05508-900, Brazil\\
$^{2}$Universidade Estadual de Santa Cruz, Rodovia Jorge Amado km 16, Ilh\'eus 45662-000, Brazil\\
$^{3}$Dipartimento di Fisica e Astronomia `Galileo Galilei', Universit\`a
di Padova, Vicolo dell'Osservatorio 3, Padova, I-35122, Italy\\
$^{4}$Universidade Federal do Rio Grande do Sul, Departamento de Astronomia, CP 15051, Porto Alegre 91501-970, Brazil
}

\date{Accepted XXX. Received YYY; in original form ZZZ}

\pubyear{2019}

\begin{document}
\label{firstpage}
\pagerange{\pageref{firstpage}--\pageref{lastpage}}
\maketitle

\begin{abstract}
Globular clusters (GCs) are the most ancient stellar systems in the Milky Way. Therefore, they play a key role in the understanding of the early chemical and dynamical evolution of our Galaxy. Around 40\% of them are placed within $\sim4$ kpc from the Galactic center. In that region, all Galactic components overlap, making their disentanglement a challenging task. With Gaia DR2, we have accurate absolute proper motions for the entire sample of known GCs that have been associated with the bulge/bar region. Combining them with distances, from RR Lyrae when available, as well as radial velocities from spectroscopy, we can perform an orbital analysis of the sample, employing a steady Galactic potential with a bar. We applied a clustering algorithm to the orbital parameters  apogalactic distance and the maximum vertical excursion from the plane, in order to identify the clusters that have high probability to belong to the bulge/bar, thick disk, inner halo, or outer halo component. We found that $\sim 30\%$ of the clusters classified as bulge GCs based on their location are just passing by the inner Galaxy, they appear to belong to the inner halo or thick disk component, instead. Most of GCs that are confirmed to be bulge GCs are not following the bar structure and are older than the epoch of the bar formation.
\end{abstract}




\section{Introduction}\label{sec:intro}

The orbital evolution of globular clusters (GCs) in the Galaxy, combined with kinematics and stellar population analyses, can provide important information to decipher the history of our Galaxy. In \citet{Bica+2016} 43 clusters were selected as related to the bulge component based on their location. Recently, \citet{Bica+19} compiled 200 GCs in the Galaxy, adding new entries such as the recently detected ones in the Vista Variables in the Via Lactea \citep{Saito+12} survey by \citet{Minniti+17a,Minniti+17b}. By
including confirmed GCs plus candidates, the number rises to 294 objects, thus reducing the lack of the Galactic GCs with respect to M31 \citep{Caldwell+Romanonwsky16}.

The stellar population components in the Galaxy have been under scrutiny in the last decades.
The Galactic disk and halo are well established, and the existence of a thick disk is increasingly more confirmed. As for the Galactic bulge configuration as an entity, it still is under definition. 
In early work, a sample of metal-rich GCs with a flattened distribution were identified as disk GCs by \citet{Zinn80,Zinn85}. The properties of this disk system of 
metal-rich GCs were described by \citet{Armandroff89}, to have a scale height between $0.8$ to $1.5$ kpc, a rotational velocity of
$193\pm29$ km s$^{-1}$, and a line-of-sight velocity dispersion of $59\pm14$ km s$^{-1}$, concluding that these parameters are consistent with properties of thick-disk stars. Later, \citet{Frenk+White82}, \citet{Minniti95}, and \citet{Cote99} deduced that metal-rich GCs within 3 kpc
from the Galactic center, should be associated to the Galactic bulge rather than to the thick disk, based on kinematics, spatial distribution and metallicity.

The Galactic bulge formation is under study in several contexts. \citet{Barbuy+18a} list the following possible scenarios: a) bulge and thick disk were formed early on simultaneously  through strong gas accretion;  
b)  hierarchical merging of subclumps; c) merging of early thick-disk subclumps; d) a major merger; e) secular evolution of the bar; f) dwarf galaxies accretion. There is also the hypothesis of the oldest stellar populations in the Galactic bulge to be seen as an extension of the inner halo in the innermost Galaxy.

The dynamical properties and their orbits provide further information that would
help allocating the objects to a given stellar population.
With the recently available Gaia Data Release 2 \citep[DR2;][]{Gaia18a},
the investigation of cluster orbits is made timely.
Previously this kind of work was attempted
for the bulge clusters, by \citet{Dinescu+03}.
More recently \citet{Perez-Villegas+18} and \citet{Rossi+15b} derived proper motions and orbits for 9 inner bulge clusters.

In the present work we study a sample of bulge GCs,
plus the intruders or outsiders from
\citet[their Tables 1 and 2]{Bica+2016}.
In order to compute orbits, reliable
radial velocities, proper motions, and distances are needed. 
Radial velocities, that
can only be derived from spectroscopy, are still missing for about a dozen of
these clusters, making that 78 of them can be studied.
The dynamical parameters perigalactic and apogalactic distances, maximum height from the plane and eccentricity, are derived such that their appartenance to different Galaxy components are identified.

In Sect. \ref{data} the sample and available data are described. In Sect. \ref{sec:model} the Galactic model potential is detailed. In Sect. \ref{sec:properties} dynamical properties of the orbits are discussed, and in Sect. \ref{sec:classi} the globular clusters are classified. The bulge globular clusters are further discussed in Sect. \ref{sec:BulgeGC}, and conclusions are drawn in Sect \ref{sec:conclusions}.

\section{Data}\label{data}

\begin{figure}
	\includegraphics[width=\columnwidth]{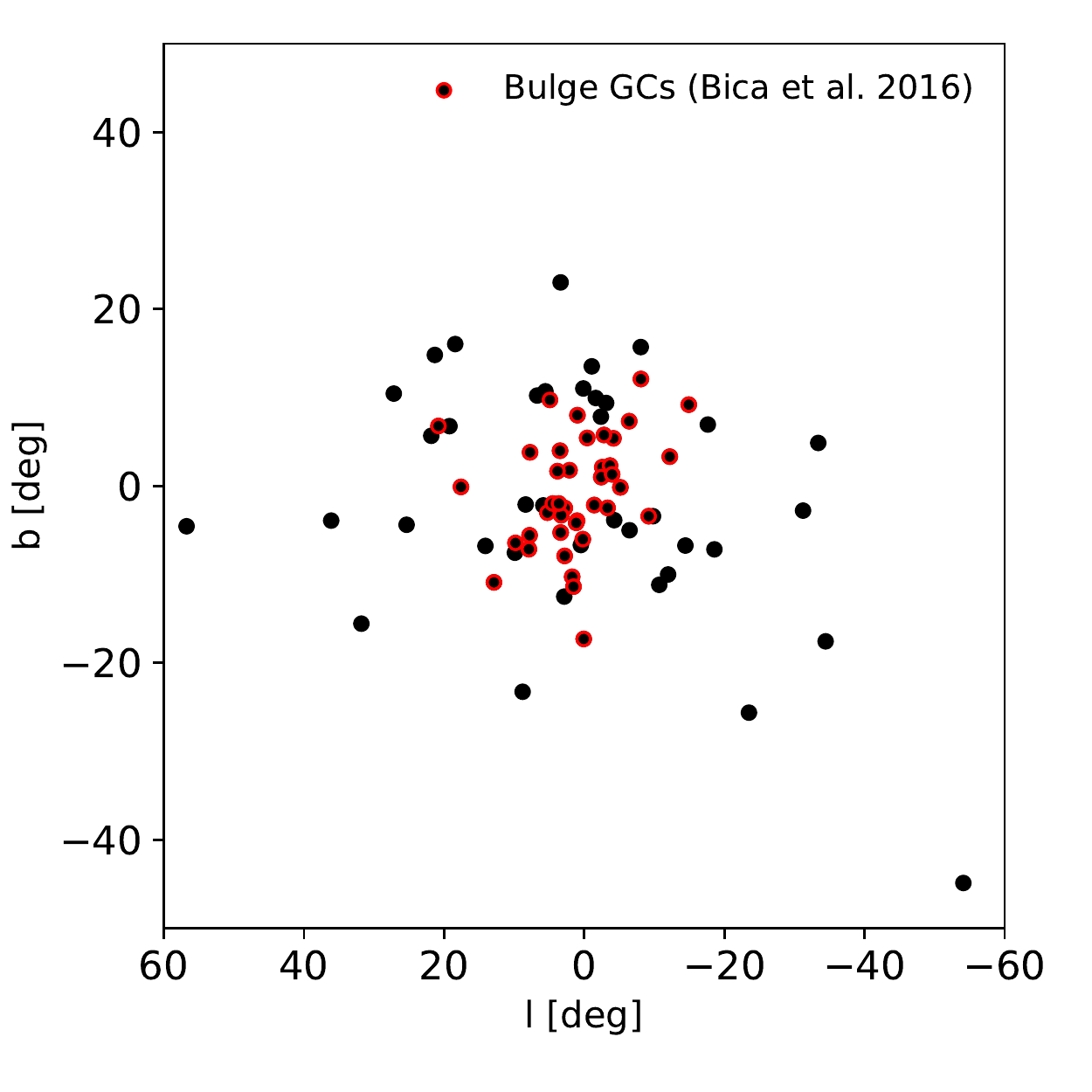}
    \caption{Location in Galactic coordinates of the GCs analyzed. Red open circles are the clusters classified as bulge GCs in \citet{Bica+2016}.
}
    \label{fig:sample}
\end{figure}

The sample includes 40 GCsfrom the list of bulge globular clusters
by \citet[their Table 1]{Bica+2016}, for which radial velocities are available. Another 37 GCs from the outer shell and clusters somewhat farther,
at distances  3.0 $<$ d$_{\rm GC}<4.5$ kpc and some clusters with d$_{\rm GC}>4.5$ kpc from the Galactic center \citep[their Table 2]{Bica+2016}, and the halo GC NGC 6752 are also added, with the purpose of identifying differences in the orbital properties among clusters that belong to each Galactic component. Figure \ref{fig:sample} shows the location in Galactic coordinates of our total sample that contains 78 clusters.

In Table \ref{tab:GC_data}, we list the cluster parameters employed in this study. Equatorial $(\alpha,\delta)$ and Galactic $(l,b)$ coordinates are given in Columns 2 to 5. The heliocentric distance $d_\odot$, in Column 6 is taken from Colour-Magnitude Diagrams (CMDs) and RR Lyrae star distance determination for the clusters HP 1, Terzan 10, NGC 6522, NGC 6626, NGC 6558, Djorgovski 1, Djorgovski 2, NGC 6304, NGC 6624, NGC 6637, NGC 6652, NGC 6717, NGC 6723, and NGC 6362  \citep[][Oliveira et al. in prep.]{Kerber+18, Barbuy+18b, Kerber+19,Ortolani+19a,Ortolani+19b} that we consider reliable. For the others, distances from \citet[][edition of 2010, hereafter H10]{Harris96}, or \citet{Baumgardt+19} were adopted. Radial velocities $V_r$, given in Column 7, are taken from spectroscopy analyses from the literature. The absolute proper motions (PMs), $\mu_\alpha^*=\mu_\alpha {\rm cos} \delta$, $\mu_\delta$, in Column 8 and 9, are the values estimated from the Gaia DR2 \citep{gaiab} and for the clusters marked with a star we calculated the PMs following the procedure explained below, the uncertainties of PMs include the systematic error of $0.035$ mas yr$^{-1}$ reported by the \citet{gaiab}. The metallicties, given in Column 10, available from high-resolution studies reported in \citet[their Table 3]{Bica+2016}, \citet{Barbuy+14,Barbuy+18b} for NGC 6522 and NGC 6558, derivations from literature and isochrone fittings to Colour-Magnitude Diagrams from our group
(Oliveira et al. 2019, in preparation),  and for the remaining clusters the values are adopted from the website by Bruno Dias
\footnote{http://www.sc.eso.org/$\sim$bdias/catalogues.html}. Column 11 gives the core radius taken from H10.

\begin{figure*}
	\includegraphics[width=2.1\columnwidth]{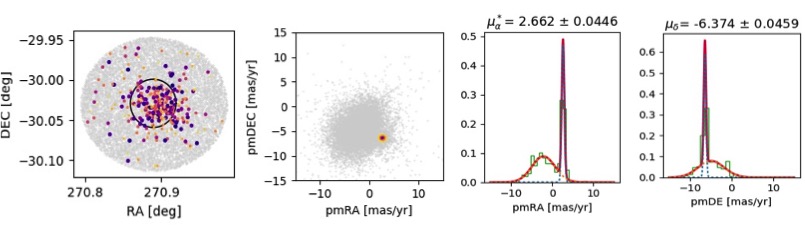}
    \caption{Example of proper motion data from Gaia DR2 for NGC 6522, and cluster member selection. First panel: core radius containing the field studied; second panel: location of the cluster in the vector-point diagram; third and fourth panels: Gaussian of RA and DEC proper motions respectively.
}
    \label{fig:PMs}
\end{figure*}

For the proper motion derivation from Gaia DR2, we selected the individual stars with the astrometric information including positions and PMs within 5 arcmin from the cluster center. We removed the stars with PM errors $>$ 0.25 mas yr$^{-1}$. Then, stars in each cluster were retrieved, within the core radius of each cluster from H10 to estimate the PMs. A selection of member stars was carried out with a combined plot of PMs in both directions, and through a Gaussian mixture model, as shown in Fig. \ref{fig:PMs}, in order to get the mean PM in each direction and uncertainties. We applied this method to 32 GCs of our sample, and the PM determinations for them are inside 1$-\sigma$ compared with those estimated by \citet{gaiab}, \citet{Vasiliev19}, and \citet{Baumgardt+19}, expet for Terzan 1 and Terzan 5, that are inside 3$-\sigma$. The PMs for the remaining clusters are taken from the new PM determination of  GCs using Gaia DR2 from the references above.

\input{TData_GCs-2.tex}
\input{TDistance_GCs-2.tex}
\twocolumn

In the past, the measurement of absolute PMs was a difficult task, especially for GCs in the innermost Galaxy, and when available, the uncertainties were large,
whereas now with Gaia that issue is essentially solved. Presently, the input distance values are the major source of uncertainty to construct more precise orbits along the Galaxy, and therefore, in the classification of the GCs into a specific stellar population. The distance values for some of them, in particular for the most reddened ones, can be significantly different from  H10  or \citet{Baumgardt+19} with their kinematic distance estimations.

Table \ref{tab:GC_dist} shows a compilation of distances from the literature. The distances used for the orbital integration is given in column 2, the distance from H10 catalog is in column 3, the kinematics distances by \citet{Baumgardt+19} in column 4, distance determinations by \citet{Valenti+07, Valenti+10} in column 5, the distance estimated by the inverse of the parallax from \citet{gaiab} is in column 6, distances taken from isochrone fittings of the CMD and RR Lyrae estimated by our group is given in column 7, and in column 8 is given the distance compilation by \citet{Bica+06}. 

The GCs with the largest difference in distance determinations in the literature are Djorg 1, Djorg 2, ESO452SC11, NGC 5927, NGC 6235, NGC 6256, NGC 6316, NGC 6380, NGC 6388, NGC 6401, NGC 6540, NGC 6553, NGC 6638, Palomar 6, Terzan 4, Terzan 5, and Terzan 9. 

In order to have a reliable statistical estimate of the uncertainty in distances, we first compared distances from  \citet{Bica+06} with those from \citet{Valenti+07,Valenti+10} for 17 bulge clusters in common. 
 These two samples are independent because the first is
 mostly optical-based
 and the other is near-infrared-based, with different reddening dependence and
 different isochrone sets.
 From the whole sample we have an uncertainty $\sigma$ = 1.3 kpc,
 but by removing two
 outliers with errors larger than 2 kpc, which clearly
 have some specific problems, the uncertainty goes down to
 $\sigma$ = 0.99 kpc. 
A second check comparing our adopted distances with respect to
\citet{Baumgardt+19} and removing two outliers with differences
larger than 4 kpc, this results in a standard deviation of 0.4 kpc.
From these two comparisons, we adopt a mean standard deviation 0.7 kpc.

Note that \citet{Hilker+19} report an error on their proper
motion distances of 8\% at the distance of 7 kpc,
 which gives an error of 0.6 kpc  at the typical
distance of the bulge clusters of 7 kpc.
This is very consistent with the suggested 0.7 kpc standard deviation.

\section {The Galactic Potential}
\label{sec:model} 

\begin{figure}
	\includegraphics[width=\columnwidth]{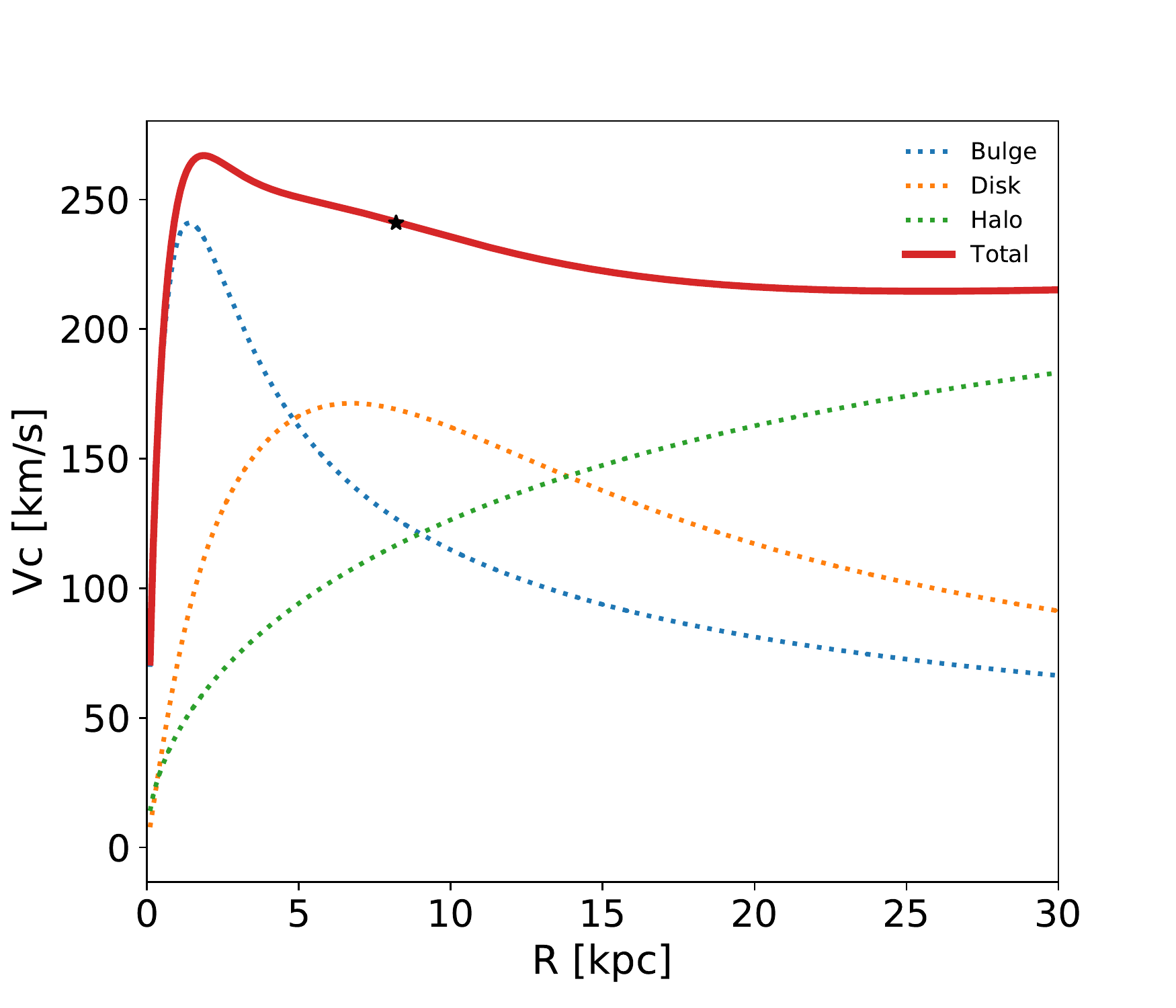}
    \caption{Total circular velocity curve and the contribution of each component of axisymmetric Galactic mass model. The black star shows the velocity at the Sun position.}
    \label{fig:RotCurve}
\end{figure}

In order to construct the orbits of the GCs, in this study we employ a non-axisymmetric model for the Galactic gravitational potential, that is built from an axisymmetric background that includes a S\'ersic bulge, an exponential disc made by the superposition of three Miyamoto-Nagai potentials \citep{MN75}, following the recipe given by \citet{Smith+15}, and a Navarro-Frenk-White (NFW) density profile \citep{NFW97} to model the dark matter halo, having a circular velocity $V_0=241$ km s$^{-1}$ at $R_0=8.2$ kpc \citep{Bland-Hawthorn16}. Even though the mass of the dark matter halo is twice larger than usual halo masses of $1.0 -1.5 \times 10^{12}$ $M_\odot$, with the scale radius we assumed, the rotation curve inside 30 kpc of our model is in agreement with \citet{Bland-Hawthorn16}, as shown in Figure \ref{fig:RotCurve}. We also adopted a less massive DM halo, also reducing the scale radius, and not significant differences are found.  

For the Galactic bar, we used a triaxial Ferrer's ellipsoid, where all the mass from the bulge component is converted into a bar. For the bar potential, we consider a total bar mass of $1.2 \times 10^{10}$ M$_{\odot}$, an angle of $25^{\circ}$ with the Sun-major axis of the bar, a major axis extension of $3.5$ kpc, and a gradient of the bar pattern speed values assumed to be $\Omega_b= 40$, $45$, and 50 km s$^{-1}$ kpc$^{-1}$. We keep the same bar extension in all cases, for any of the bar pattern speed values. Table \ref{tab:parameters} gives the parameters used for our Galactic model.

\begin{table}
\caption{Parameters of the adopted Galactic mass model.}
\label{tab:parameters}
\begin{tabular}{l c c}
\hline
\\
Parameter&Value & Reference\\
\hline
\multicolumn{3}{c}{\textbf{Axisymmetric components}}\\
\hline
\multicolumn{3}{c}{\textbf{Bulge}}\\  
 $M_\mathrm{b,tot}$ &$1.2 \times 10^{10}$ $M_\odot $ & 1, 2 \\
 $R_\mathrm{e}$ & $0.87$ $\mathrm{kpc}$ &  \\
 $n$ &$1$ & 3\\
\\
\multicolumn{3}{c}{\textbf{Disc}}\\
$M_\mathrm{d,1}$ & $1.07 \times 10^{11}$ $M_\odot $ & \\
$a_\mathrm{d,1} $ &$5.97$ $\mathrm{kpc}$ &\\
$M_\mathrm{d,2}$ & $-7.09 \times 10^{10}$ $M_\odot $ & \\
$a_\mathrm{d,2} $ &$12.99$ $\mathrm{kpc}$ &\\
$M_\mathrm{d,3}$& $1.2 \times 10^{10}$ $M_\odot $ & \\
$a_\mathrm{d,3} $ & $2.04 \;\mathrm{kpc}$ &\\
$b_\mathrm{d}$ & $0.25\; \mathrm{kpc}$ &\\               
\\
\multicolumn{3}{c}{\textbf{Halo}}\\
$M_\mathrm{h} $ & $3.0 \times 10^{12} \; M_\odot $ &  4\\ 
$a_\mathrm{h}$ & $57.0 \; \mathrm{kpc}$ & 4\\
\\
\hline
\multicolumn{3}{c}{\textbf{Galactic bar}}\\
\hline
 $M_\mathrm{bar}$ & $1.2 \times 10^{10}$ $M_\odot $ & 1, 2, 5\\
 $n$ & $2$ & 6\\
 $a$ & $3.5 \; \mathrm{kpc}$ &9, 10 \\
 $b$ & $1.4 \; \mathrm{kpc}$ & 9, 10\\
 $c $ &$1.0 \; \mathrm{kpc}$ &9, 10 \\
 $\phi_\mathrm{bar}$ & $25^\circ$  & 2,7, 8 \\ 
 $\Omega_\mathrm{bar} $ &$40, 45, 50$ $\mathrm{km} \,\mathrm{s}^{-1}\, \mathrm{kpc}^{-1}$ &2, 11, 12, 13 \\
\hline
\hline
\end{tabular}
\\
\\
{\footnotesize{ \textbf{References.} (1) \citealt{Portail+15a}; (2) \citealt{Bland-Hawthorn16}; (3) \citealt{Kent+91}; (4) \citealt{Irrgang+13}; (5) \citealt{Weiner99}; (6) \citealt{Pfenniger84}; (7) \citealt{Rattenbury+07}; (8) \citealt{Wegg+13}; (9) \citealt{Freudenreich98}; (10) \citealt{Gardner+10}; (11) \citealt{Bissantz+03}; (12) \citealt{Portail+17}; (13) \citealt{Perez-Villegas+17}.}}
\end{table}

\section{Properties of the Globular Cluster Orbits}\label{sec:properties}

To construct the GC orbits through the Galactic potential, we have employed the Shampine--Gordon integration scheme that is implemented in the \textsc{nigo} tool \citep[Numerical Integrator of Galactic Orbits --][]{Rossi15a}. 

We integrated the orbits for the present cluster sample forward in time for 10 Gyr. Using the observational parameters of the GCs given in Table \ref{tab:GC_data}, we computed the initial state vector of the clusters assuming the Sun's Galactocentric distance $R_0=8.2$ kpc and a circular velocity $V_0=241$ km s$^{-1}$ \citep[e.g.][and references therein]{Bland-Hawthorn16}. The velocity components of the Sun with respect to the local standard of rest (LSR) are $(U,V,W)_{\odot}= (11.1, 12.24, 7.25)$ km s$^{-1}$ \citep{Schonrich+10}. The velocity components of the cluster in the heliocentric reference system $U,V,$ and $W$ are positive in direction of the Galactic center, Galactic rotation, and North Galactic Pole, respectively. We defined as inertial Galactocentric frame of reference the right--handed system of coordinates ($x,y,z$) where the $x$--axis points to the Sun from the Galactic Centre and the $z$--axis points to the North Galactic Pole. We refer to the bar--corotating frame of reference as the right--handed system of coordinates ($x_\mathrm{r},y_\mathrm{r},z_\mathrm{r}$) that co--rotates with the bar, where the $x_\mathrm{r}$ axis is aligned with the bar semi--major axis $a$ and $z_\mathrm{r}$ points towards the North Galactic Pole.

\begin{figure}
	\includegraphics[width=1.1\columnwidth, height=1.5\columnwidth]{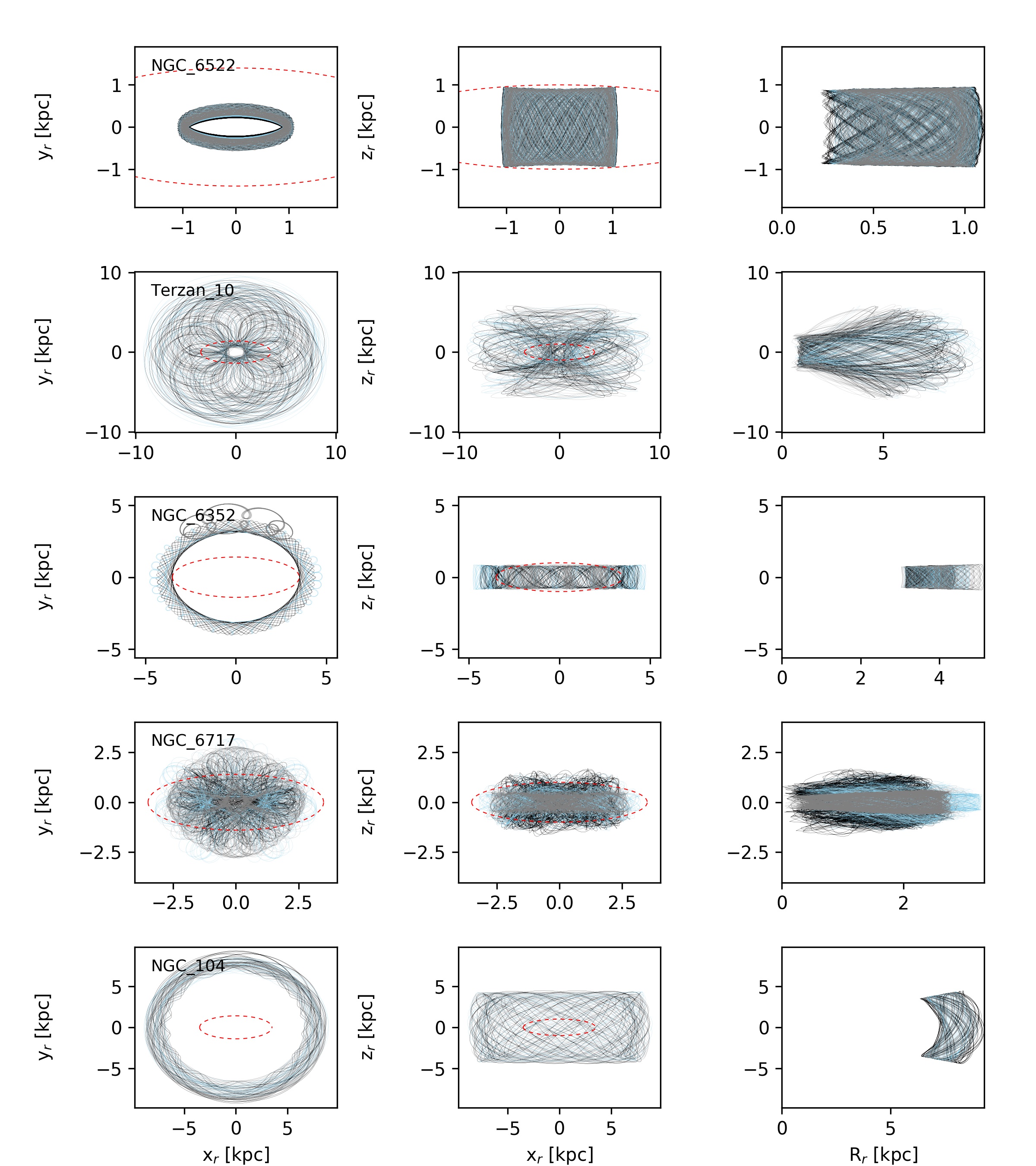}
    \caption{Examples of orbits for the globular clusters. The three columns show $x-y$, $x-z$, and $R-z$ projections for orbits with the non-axisymmetric Galactic potential co-rotating with the bar. The colors in the left panels are the orbits with different pattern speed of the bar, 40 (black), 45 (blue), and 50 (grey) km s$^{-1}$ kpc$^{-1}$. The dashed red line shows the size of the Galactic bar.
}
    \label{fig:exa_orbits}
\end{figure}

For each cluster we generate a set of 1000 initial conditions, where we follow the Monte Carlo technique taking into account the uncertainties on 
the heliocentric distance, PMs in both directions, and the radial velocity given in 
Table \ref{tab:GC_data}. The PM uncertainties include the average of the systematic error of $0.035$ mas yr$^{-1}$ reported by the \citet{gaiab}.
These calculations were carried out with the purpose of evaluating how much the orbital properties of the GCs change due to the uncertainties of the observed data. Figure \ref{fig:exa_orbits} shows some examples of orbits with the three different bar pattern speed values, in the frame co-rotating with the bar, for five GCs of our sample: NGC 6522, Terzan 10, NGC 6352, NGC 6717, and NGC 104 (from top to bottom). The initial conditions for them are the central values given in Table \ref{tab:GC_data}.

The orbital properties we calculate, in the inertial frame of reference, are the perigalactic distance $r_{\mathrm{min}}$, the apogalactic distance $r_{\mathrm{max}}$, the maximum vertical excursion from the Galactic plane $|z|_{\mathrm{max}}$, and the eccentricity defined as $e=(r_{\mathrm{max}}- r_{\mathrm{min}} )/(r_{\mathrm{max}}+ r_{\mathrm{min}} )$.

In Table \ref{tab:dynP}, we present the orbital parameters of our sample of GCs, where we use three values of
rotational velocity of the bar $\Omega_b= 40$, $45$, and $50$ km s$^{-1}$ kpc$^{-1}$. For each bar pattern speed, we give the average values of the perigalactic distance, apogalactic distance, the maximum vertical excursion from the Galactic plane, and the orbital eccentricity. The errors provided in each column are the standard deviation of the distribution. We see that for the clusters that are confined in the innermost Galaxy, such as HP 1, Djorg 2, Liller 1, and NGC 6325, the effect of the bar pattern speed is almost negligible. On the contrary, for NGC 5927, the perigalactic distance decreases with the pattern speed.

\input{TOrbital_para-2.tex}

\clearpage
\onecolumn
\begin{figure*}
	\includegraphics[width=1.1\columnwidth]{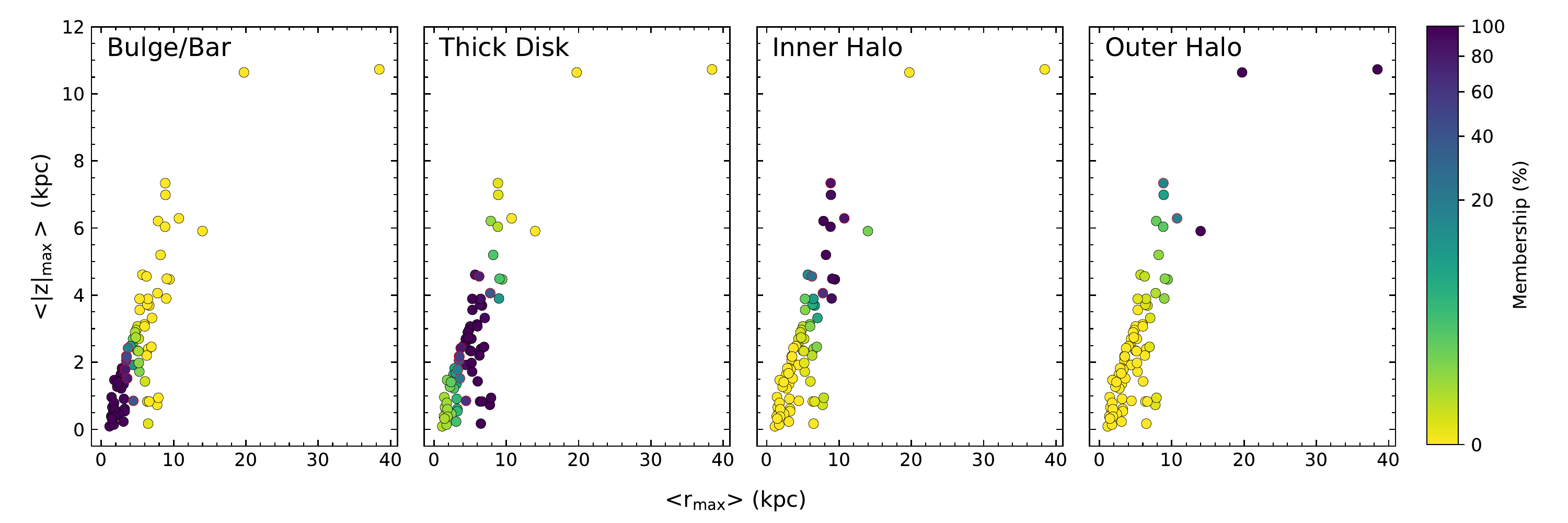}
    \caption{Membership probability for $\Omega_b=$ 40 km s$^{-1}$ kpc$^{-1}$. From left to right panel shows the probability that each GC has to belong to the bulge/bar, thick disk, inner halo, and outer halo, respectively. Open red circles highlight the clusters with probabilities between 15\% to 85\% that are in the boundary of two Galactic component.
}
    \label{fig:Membership}
\end{figure*}

\input{TMembership.tex}

\section{Classifying globular clusters}\label{sec:classi}

In the literature, GCs have been classified based on their spatial distribution, metallicity, and internal and structural parameters 
\cite[e.g.][]{Aguilar+88,Minniti95,Barbuy+98,Cote99,Bica+2016,Pasquato+19}. In this section, we used the orbital parameters of the 78 GCs, calculated in Section \ref{sec:properties}, to separate the clusters into different stellar populations. We obtained this classification using unsupervised clustering algorithms.

\subsection{Clustering method}\label{sec:clust}

The identification of GCs that are confined or belong to the bulge/bar component, which is the main aim of the present work, is not an easy task because in the innermost part of the Galaxy, all components overlap: 
 the bulge/bar, thick disk, inner halo, and outer halo. 
 
 In terms of dynamics, the GCs of each component should share its dynamical properties. Based on this criterion, we employed the orbital parameters (Table \ref{tab:dynP}) to separate the GCs into four Galactic components: bulge/bar, thick disk, inner halo, and outer halo. To do this classification, we employ the Gaussian Mixture
Models (GMM), an unsupervised machine learning algorithm 
that searches for 
$K$ Gaussian distributions, which fits better a $N$D parameter space. Based on 
Bayes's theorem, the GMM algorithm tries to maximize the expression:

\begin{center}
 $ \displaystyle G(x) = \sum_{i=1}^{K} \phi_i \times \mathcal{N}\left( x | \mu_i, \sigma_i \right), $
\end{center}

where $\mathcal{N}( x | \mu_i, \sigma_i )$  represents the ith Gaussian distribution 
with mean $\mu_i$ and standard deviation $\sigma_i$. The $\vec{x}$ is the parameter space.

We employed the 2D GMM clustering algorithm using two orbital parameters: the apogalactic distance $r_{max}$ and the maximum height from the plane $|z|_{max}$. For the GCs separation, we take into account the contribution of the set of orbits of each GC and the three bar pattern speeds. To prevent confusion in the clustering method, we removed the orbits with $r_{max} > 20$ kpc.

From the GMM algorithm, we obtain the centre $\mu_i (r_{max},|z|_{max})$, width $\sigma_i (r_{max},|z|_{max})$, and weight $P_i$ of each stellar component. For the bulge/bar component $\mu_{b}= (2.50, 1.07)$ kpc, $\sigma_{b}=(0.83, 0.63)$ kpc, and $P_{b}$=0.38; for the thick disk $\mu_{d}= (5.60, 2.42)$ kpc, $\sigma_{d}=(1.29, 1.09)$ kpc, and $P_{d}$=0.46; for the inner halo $\mu_{H_i}= (8.60, 5.12)$ kpc, $\sigma_{H_i}= (1.31, 1.02)$ kpc, and $P_{H_i}$=0.11, and outer halo $\mu_{H_o}= (12.75, 8.47)$ kpc, $\sigma_{H_o}= (2.57, 2.18)$ kpc, and $P_{H_o}$=0.032.

For the present work, a statistical approach was necessary since the classification 
depends on the distribution of the data in the parameter space and their uncertainties. 
The GMM was applied from the library \texttt{scikit-learn} \citep{Pedregosa+11}.

\subsection{Membership probability and GC classification}

With the information of the each component provided by the GMM and using a Gaussian distribution probability, we calculate the membership probability that each cluster has to belong to each Galactic component. Figure \ref{fig:Membership} shows the membership probability for $\Omega_b=$ 40 km s$^{-1}$ kpc$^{-1}$. The dark blue color indicates the clusters that have probability higher than 85\% to belong to the  bulge/bar (first panel), thick disk (second panel), inner halo (third panel), and outer halo (four panel) component. The red open circles mark the clusters with probability between 15\% to 85\%, those clusters are in the boundary between two Galactic components. Table \ref{tab:member_prob} gives the probability for each cluster to be part of the bulge, disk, inner halo, and outer halo, for the three pattern speeds of the bar.

\begin{figure*}
	\includegraphics[width=1.9\columnwidth]{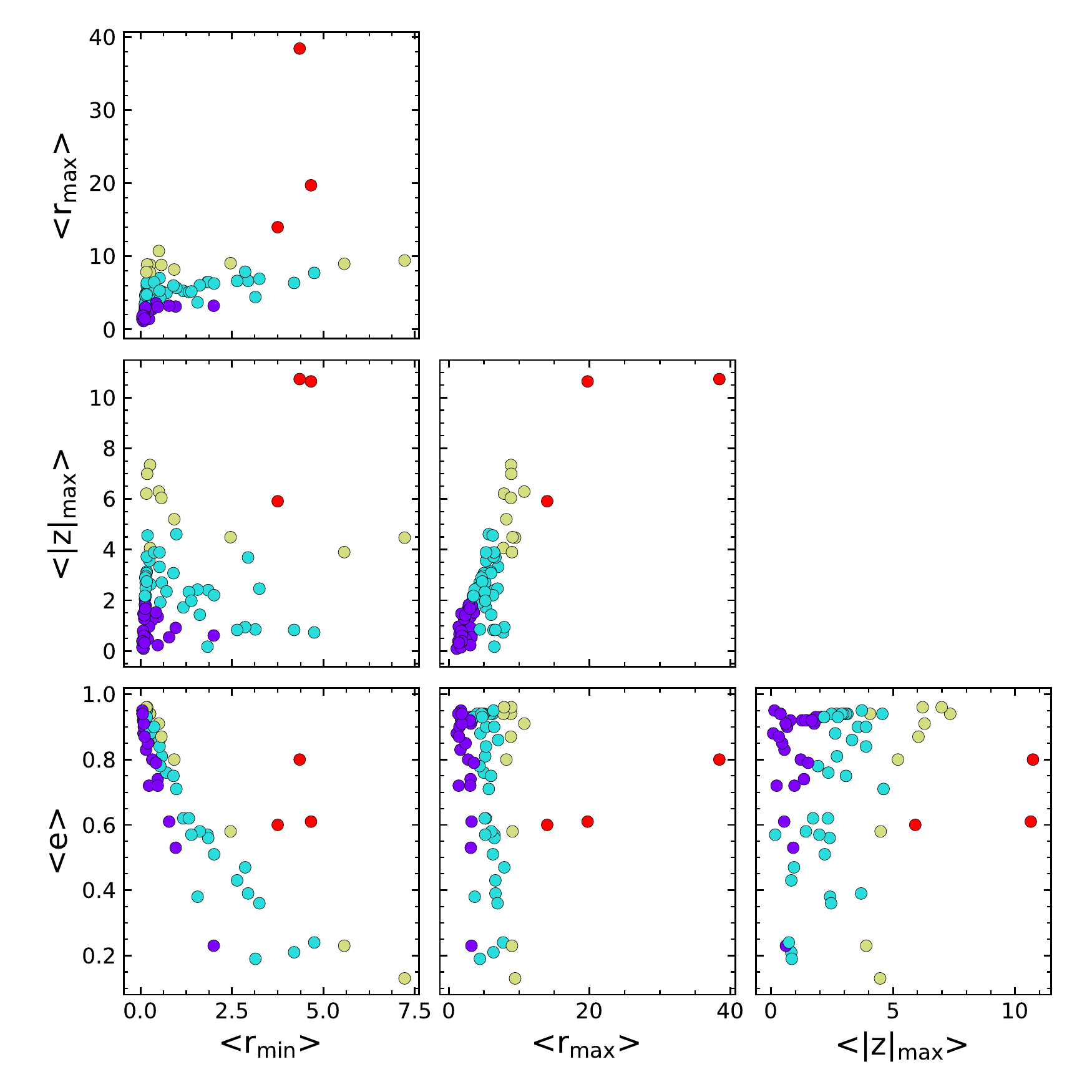}
    \caption{Orbital parameters as function of the median value ot the perigalactic distance $<r_{min}>$,  the apogalctic ditance $<r_{max}>$, the maximum distance from the Galactic plane $<|z|_{max}>$, and average eccentricity $e$, for $\Omega_b=$ 40 km s$^{-1}$ kpc$^{-1}$. The four groups are: bulge/bar (purple), thick disk (cyan), inner halo (green), and outer halo (red).
}
    \label{fig:clustering}
\end{figure*}

The maximum probability of membership given in Table \ref{tab:member_prob} was used to classify the GCs into each Galactic component. Figure \ref{fig:clustering} shows a plot of combinations among the orbital parameters, where the colours identify the groups of clusters based on the membership probability.  
Figure \ref{fig:clustering} also shows a clear correlation between $<r_{max}>$ and $<|z|_{max}>$, and an anticorrelation between $<r_{min}>$ and eccentricity. 

In Table \ref{tab:classification}, we list the GCs that belong to each of the Galactic components. Our sample has 40 out of 43 GCs identified as bulge GCs in \citet{Bica+2016}, and based on their dynamical properties, for the pattern speed of 40 km s$^{-1}$ kpc$^{-1}$, we found that 27 of them, have characteristics of bulge GCs with a higher probability, whereas the other 13 GCs (Mercer 5, NGC 6316, NGC 6325, NGC 6355, NGC 6539, NGC 6638, NGC 6652, Ton 2, Palomar 6, Djorg 1, Terzan 3, Terzan 10, and NGC 6723) appear to be intruders from other stellar components, that currently are crossing the central parts of the Galaxy, provided that
their distances are confirmed in the future. Additionally, the cluster NGC 6380 and NGC 6569 are part of the bulge GCs in our clasification, therefore we have 29 bulge GCs. Another 37, 9, and 3 GCs are identified as thick disk, inner halo, and outer halo, respectively. There are three clusters, NGC 6535, NGC 6284, and NGC 6333, that change of component with the bar pattern speed. 
Additionally, there are clusters with significant probability to be in two stellar components.  ESO452SC11, NGC 6325, NGC 6352, NGC 6539, NGC 6569, NGC 6626, and NGC 6638 are in the boundary between bulge/bar and thick disk; NGC 6273 and NGC 6535 are between thick disk and inner halo; NGC 6284 and NGC 6333 are between inner halo and outer halo,  and those clusters could change with the bar pattern speed. It is important to stress that due to the uncertainties of the data, in particular in the distances, the classification established here, might change in the future when
better distance determinations are available.
One example is Palomar 6, that with the distance of d$_{\odot}$ = 5.8 kpc puts it in the thick disk component whereas with the distance by \citet{Ortolani+95}
of d$_{\odot}$ = 8.9 kpc puts it in the bulge.

\begin{figure*}
	\includegraphics[width=2\columnwidth]{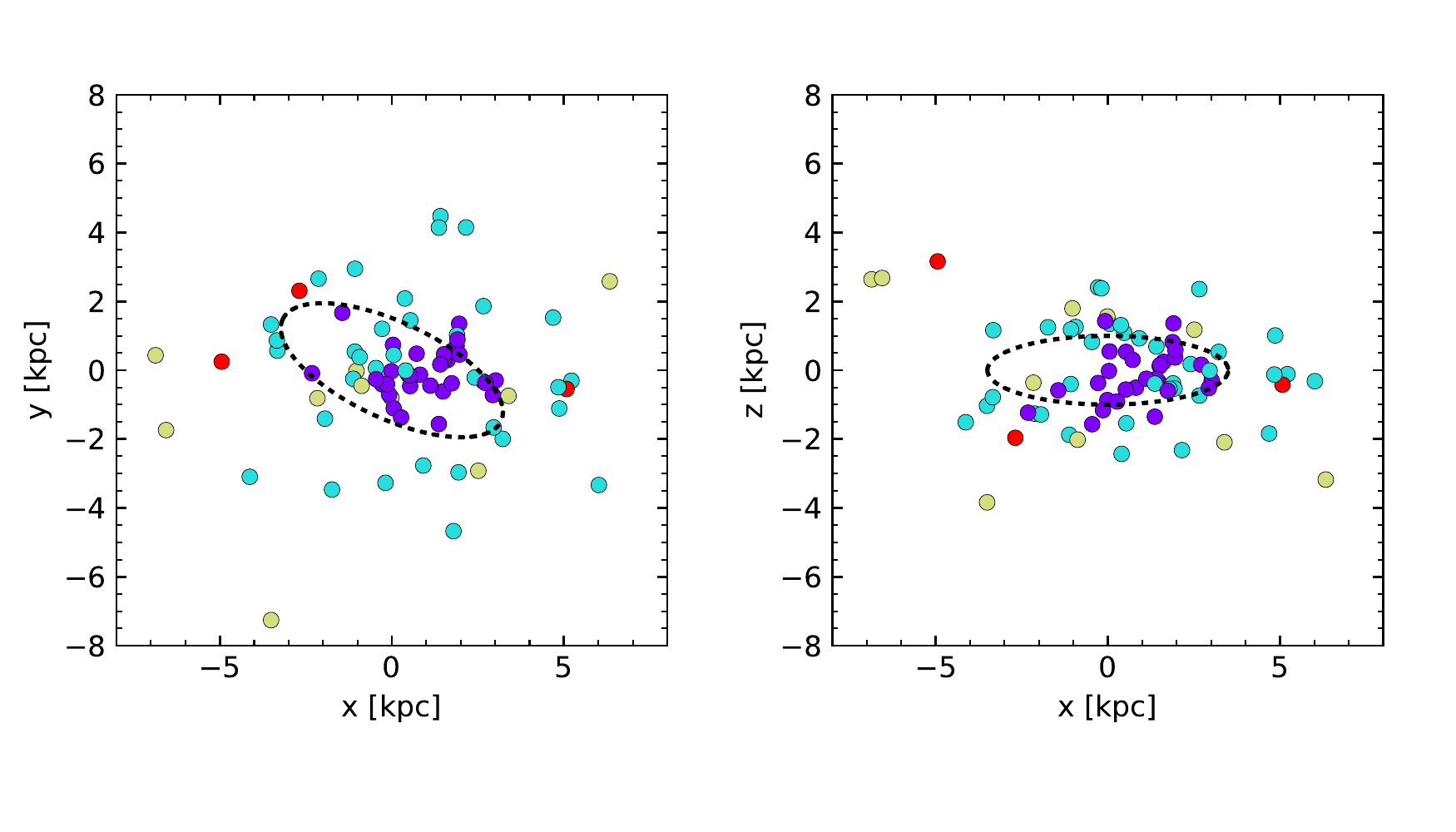}
    \caption{Spatial distribution of the GCs in the $x-y$ (left panel) and $x-z$ (right planel) projection. The colours are the same as in Figure \ref{fig:clustering} and indicate the stellar component to which GCs are associated. The  black dashed line shows the size of the Galactic bar.
}
    \label{fig:cluster_pos}
\end{figure*}

Figure \ref{fig:cluster_pos} shows the spatial distribution of the sample GCs, based on our classification and the stellar component with higher membership probability, 
where we can note that GCs from different Galactic components overlap in the innermost part of the Galaxy.

\input{TGCs_classification.tex}

\section{Bulge Globular clusters}\label{sec:BulgeGC}

Once we identify clusters with high probability to be bulge GCs, we analyze their set of orbits, with the purpose of knowing which bulge GCs follow the bar. To classify orbits, we use the same criteria as \citet{Portail+15b}. The orbit classification employs frequency analysis. First, we compute the fast Fourier transform for each orbit in the Cartesian $x$ coordinate and the Cylindrical radius $R$, to identify the main frequencies. Then, the orbits for which the frequency ratio $f_R/f_x = 2 \pm 0.1$ are bar-following orbits. The orbits that are not supporting the bar-shape have a frequency ratio $f_R/f_x \neq 2 \pm 0.1$. Table \ref{tab:followBar} gives the percentage of orbits in each bulge GC that support the bar shape. The clusters Liller 1, NGC 6304, NGC 6522, NGC 6528, NGC 6540, NGC 6553, Terzan 5, and Terzan 9 have more than $20\%$ of their orbits that support the bar. The fraction of orbits that follow the Galactic bar decreases with the rotation of the bar, except for the cases of NGC 6304, NGC 6342, and NGC 6637, that increases instead. Also, we found that most of the bulge/bar GCs are not suporting the Galactic bar.

\begin{table}
\caption{ Bulge GC orbits that follow the Galactic bar for different pattern speed.}
\label{tab:followBar}
\begin{center}
\begin{tabular}{l c c c}
\hline
\hline
Cluster &\multicolumn{3}{c}{Orbits following the bar}  \\
        &\multicolumn{3}{c}{(\%)} \\  
\hline
 &$\Omega_b=40$  & $\Omega_b=45$  & $\Omega_b=50$ \\
\hline
BH 261 & 6.2 & 6.9 & 8.9 \\
Djorg 2 &  0.0   &  0.1   &  0.0 \\
ESO452SC11 & 0.0 & 0.2  & 0.0    \\
HP 1 &    0.2   &  0.4   &  0.8 \\
Liller 1 &  38.2 & 32.1 & 28.1 \\
NGC 6256 &  9.0 &  19.7 & 36.1 \\
NGC 6266 &  13.1 & 9.9   &  8.3 \\
NGC 6304 &  27.8 & 38.1 & 38.4 \\
NGC 6342 &  1.0 &  3.1   &  3.5 \\
NGC 6380 &  0.7 &  0.9   &  0.3 \\
NGC 6401 &  0.7 &  0.7   &  1.4 \\
NGC 6440 &  11.3 & 9.1   &  7.5 \\
NGC 6522 &  97.0 & 95.3 & 94.7 \\
NGC 6528 &  44.0 & 24.0 &  7.8 \\
NGC 6540 &  23.1 & 18.2 & 12.9 \\
NGC 6553 &  65.2 & 63.3 & 57.0 \\
NGC 6558 &  19.3 & 12.5 &  6.9 \\
NGC 6569 &  2.0 & 2.6 &  3.1 \\
NGC 6624 &  0.3 &  0.0   &  0.1 \\
NGC 6626 &  0.4 &  0.1   &  0.5 \\
NGC 6637 &  1.3 &  4.2   &  4.4 \\
NGC 6642 &  7.2 &  2.8   &  2.1 \\
NGC 6717 &  0.7 &  0.8   &  0.7 \\
Terzan 1 &  2.0 &  2.3   &  1.6 \\
Terzan 2 &  0.0 &  0.0   &  0.0 \\
Terzan 4 &  19.6 & 15.1 & 13.7 \\
Terzan 5 &  32.8 & 25.0 & 12.3 \\
Terzan 6 &  1.7 &  2.1   &  0.5 \\ 
Terzan 9 &  87.7 & 81.8 & 73.1 \\

\hline
\hline
\end{tabular}
\end{center}
\end{table}

On the other hand, there are bulge GCs that even if they are not supporting the bar shape, they are trapped into a bar resonance such as NGC 6266 and NGC 6558, in the 3:1 resonance.

\begin{figure}
	\includegraphics[width=\columnwidth]{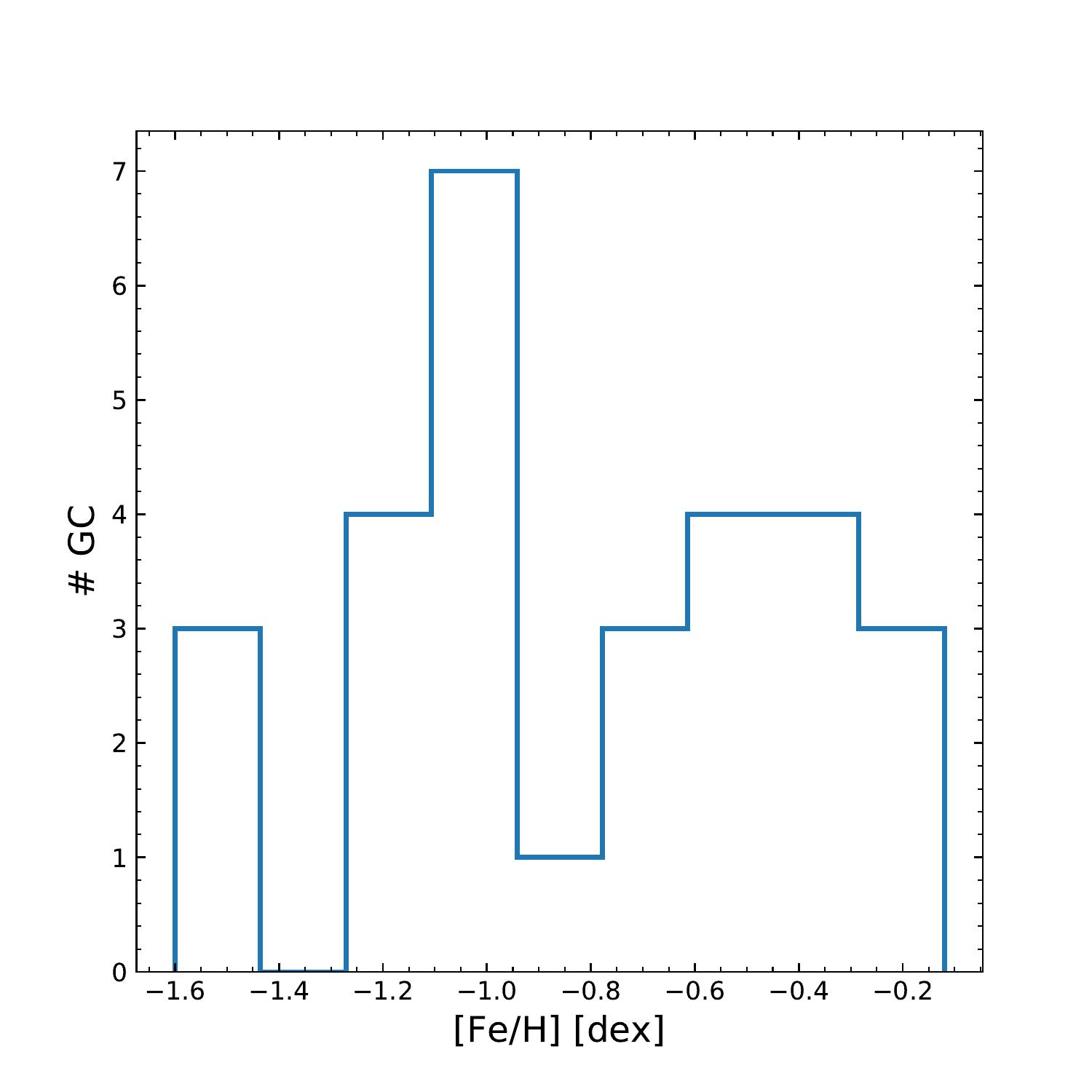}
    \caption{Metallicity distribution of bulge globular clusters.}
    \label{fig:bulge_met}
\end{figure}

\begin{figure*}
	\includegraphics[width=2.1\columnwidth]{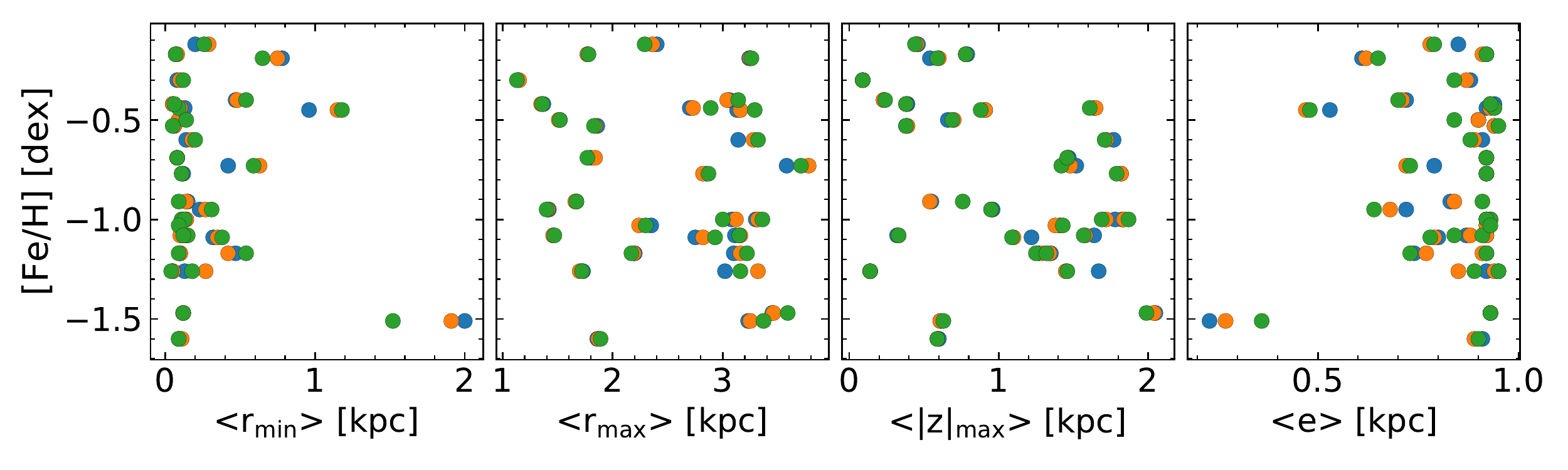}
    \caption{Metallicity of bulge globular clusters as function of the median value of perigalactic distance $<r_{min}>$ (first panel), apogalctic distance $<r_{max}>$ (second panel), maximum distance from the Galactic plane $<|z|_{max}>$ (third panel), and eccentricity $<e>$ (fourth panel), for $\Omega_b=$ 40 (blue), 45 (orange) , and 50 (green) km s$^{-1}$ kpc$^{-1}$.
}
    \label{fig:cluster_met_orbital}
\end{figure*}

Additionally, the metallicity distribution function (MDF) of the classified bulge GCs is
shown in Figure \ref{fig:bulge_met}, where we can note a clear peak at [Fe/H]$\sim-1.0$, as  previously already reported  by \citet{Rossi+15b} and \citet{Bica+2016}. Figure \ref{fig:cluster_met_orbital} gives the metallicity as a function of the orbital parameters, the perigalactic and apogalactic distances, 
the maximum height, and the eccentricity. We cannot see any trend among orbital parameters with the metallicity. Also,  the effect of changing the bar pattern speed is almost negligible for the bulge GCs.

\section{Conclusions}\label{sec:conclusions}
Combining accurate absolute PMs from Gaia DR2 with distances from RR Lyrae and/or CMD when available, and radial velocities from spectroscopy, we were able to construct the orbits of 78 GCs among the most centrally located in the Galaxy, using a Galactic bar potential. Most sample clusters are located in the innermost region of the MW.

We generate a set of initial conditions for each cluster using Monte Carlo simulations, taking into account the uncertainties from the heliocentric distance and kinematic data of the clusters. We calculated the average value of the orbital parameters such as the perigalactic and apogalactic distances, the maximum vertical excursion from the Galactic plane, and the eccentricity. Through the dynamical properties of the orbits, apogalactic distance and the maximum height from the plane, we applied a clustering technique to estimate the membership probability that each cluster has to be part of different stellar populations: bulge/bar, thick disk, inner halo, or outer halo.

Given that in the inner part of the MW, most of Galactic components overlap, one of the goals of this study was to identify which among the bulge GCs classified as such by \citet{Bica+2016}, are confined in the bulge/bar region. In a few cases, like Djorg 1 and Terzan 10, they are identified as halo intruders in the Galactic bulge \citep{Ortolani+19a}, and with our analysis, we confirm that they are in fact contaminating the bulge GC population. Others in the same case are NGC 6316, NGC 6355, Ton 2, Palomar 6, and NGC 6723, provided that their distances are confirmed. In terms of dynamical properties, the orbital characteristics of the GCs identified to be part of the bulge/bar are distinct from those in the thick disk/inner halo according to the present classification criteria.  

The GCs are very old, whereas the bar likely formed later.
In the already seminal paper by \citet{Sheth+08},
it was shown that whereas 65\% of the present-day luminous spiral galaxies have bars, at redshift z$\sim$0.84 (about 6 Gyr ago), this fraction drops to 20\%.
From a fully cosmological simulation, \citet{Buck+18} suggest that the Galaxy's bar formed 8$\pm$2 Gyr ago. 
Similarly, \citet{Bovy+19} used 
 chemical abundances from
 the {\it Apache Point Observatory Galactic Evolution Experiment} (APOGEE) survey,
 and kinematical information from the Gaia collaboration, and concluded that the Galactic bar formed $\sim 8$ Gyr ago.
Therefore, the fact that GCs are in the bar, probably indicates that they were confined in the bar when the latter formed.

Most of the bulge GCs are confined in the bar region but are not supporting the bar structure, the seven clusters that are supporting the bar do not necessarily support the X-shape. In particular Terzan 5, that is confined in the bar, was revealed to have multiple populations in a range of metallicities and ages \citep[][and references therein]{Origlia+19}, suggesting this object to be the nucleus of a dwarf galaxy.

Currently, we have very accurate proper motions from Gaia, and radial velocities from high resolution spectroscopy, however in
most cases there are no accurate distance determinations, and they are crucial to characterize the orbits. 
Further efforts on distance derivation from accurate CMDs and/or RR Lyrae, are greatly needed.

\section*{Acknowledgements}
 We acknowledge the anonymous referee for the detailed review and for the helpful suggestions, which allowed us to improve the manuscript. APV acknowledges the FAPESP postdoctoral fellowship no. 2017/15893-1. BB and EB acknowledge grants from the Brazilian agencies CAPES - Finance code 001, CNPq and FAPESP.
 SO acknowledges partial support by the Universit\`a degli Studi di Padova Progetto di Ateneo CPDA141214 and BIRD178590 and by INAF under the program PRIN-INAF2014. SOS acknowledges the FAPESP PhD fellowship no. 2018/22044-3. APV and SOS acknowledge the DGAPA-PAPIIT grant IG100319.
``This work has made use of data from the European Space Agency (ESA) mission Gaia (https://www.cosmos.esa.int/gaia), processed by the Gaia Data Processing and Analysis Consortium (DPAC, https://www.cosmos.esa.int/web/gaia/dpac/consortium). Funding for the DPAC has been provided by national institutions, in particular the institutions participating in the Gaia Multilateral Agreement.''










\bsp	
\label{lastpage}
\end{document}

%% file: TData_GCs-2.tex
\clearpage
\onecolumn
\small
\begin{longtable}{lcccccccccc}
\caption{Globular Clusters data.}
\label{tab:GC_data}

\\
\hline
   Cluster & $\alpha_{2000} $ & $\delta_{2000}$ & l & b  & $d_{\odot}$ & $V_r$ & $\mu_{\alpha}^*$ & $\mu_{\delta}$  & [Fe/H]$^\dag$ & R$_c^\ddag$\\
    & (deg) & (deg) & (deg) & (deg)& (kpc) & $(\mathrm{km}\,\mathrm{s}^{-1})$ & $(\mathrm{mas}\,\mathrm{yr}^{-1})$ & $(\mathrm{mas}\,\mathrm{yr}^{-1})$& (dex) & (arcmin) \\ \endfirsthead
\hline
\hline
BH 261    & $ 273.52$ & $ -28.63$ & $   3.36$ & $  -5.27$ & $    6.50 \pm    0.65 ^e$ & $  -29.38 \pm    0.60^e $ & $    3.59 \pm    0.05 $ & $   -3.57 \pm    0.05 $ & $  -1.17$ & $0.40$  \\ 
Djorg 1    & $ 266.86$ & $ -33.06$ & $ 356.67$ & $  -2.48$ & $    9.30 \pm    0.50^a $ & $ -358.10 \pm    0.70^i $ & $   -5.11 \pm    0.07 $ & $   -8.30 \pm    0.05 $ & $  -1.36$ & $   0.50$  \\ 
Djorg 2    & $ 270.45$ & $ -27.82$ & $   2.77$ & $  -2.50$ & $    8.75 \pm    0.12^b $ & $ -159.90 \pm    0.50^i $ & $    0.58 \pm    0.06 $ & $   -3.05 \pm    0.05 $ & $  -0.91$ & $   0.33$  \\ 
ESO452SC11 & $ 249.85$ & $ -28.39$ & $ 351.91$ & $  12.09$ & $    6.50 \pm    0.65^e $ & $   16.27 \pm    0.48^e $ & $   -1.54 \pm    0.05 $ & $   -6.41 \pm    0.05 $ & $  -1.47$ & $   0.50$  \\ 
HP 1$^\star$       & $ 262.77$ & $ -29.98$ & $ 357.42$ & $   2.12$ & $    6.59 \pm    0.16^c $ & $   40.00 \pm    0.50^j $ & $    2.41 \pm    0.06 $ & $  -10.14 \pm    0.05 $ & $  -1.00 ^m$ & $   0.03$  \\ 
Liller 1   & $ 263.35$ & $ -33.38$ & $ 354.84$ & $  -0.16$ & $    8.20 \pm    0.82^d $ & $   60.18 \pm    2.46^e $ & $   -5.53 \pm    0.52 $ & $   -7.69 \pm    0.33 $ & $  -0.30^m$ & $   0.06$  \\ 
Lynga 7    & $ 242.76$ & $ -55.31$ & $ 328.77$ & $  -2.79$ & $    8.00 \pm    0.80^e $ & $   17.86 \pm    0.83^e $ & $   -3.80 \pm    0.04 $ & $   -7.06 \pm    0.04 $ & $  -0.56$ & $   0.90$  \\ 
Mercer 5   & $ 275.83$ & $ -13.66$ & $  17.59$ & $  -0.11$ & $    5.50 \pm    0.55 ^e$ & $  185.50 \pm    3.75 ^e$ & $   -4.22 \pm    0.35 $ & $   -6.97 \pm    0.31 $ & $  -0.85$ & $   1.00$  \\ 
NGC 104    & $   6.03$ & $ -72.08$ & $ 305.89$ & $ -44.89$ & $    4.50 \pm    0.45^d $ & $  -48.00 \pm   10.00^k $ & $    5.25 \pm    0.04 $ & $   -2.52 \pm    0.04 $ & $  -0.71$ & $   0.36$  \\ 
NGC 5927   & $ 232.00$ & $ -50.68$ & $ 326.60$ & $   4.86$ & $    8.16 \pm    0.27^e $ & $  -99.00 \pm   11.00^e $ & $   -5.05 \pm    0.04 $ & $   -3.23 \pm    0.04 $ & $  -0.32$ & $   0.42$  \\ 
NGC 6139   & $ 246.92$ & $ -38.85$ & $ 342.36$ & $   6.94$ & $    9.80 \pm    0.83 ^e$ & $   24.41 \pm    0.95^e $ & $   -6.16 \pm    0.05 $ & $   -2.67 \pm    0.04 $ & $  -1.42$ & $   0.15$  \\ 
NGC 6144   & $ 246.81$ & $ -26.02$ & $ 351.93$ & $  15.70$ & $    8.90 \pm    0.89^d $ & $  195.75 \pm    0.74^e $ & $   -1.76 \pm    0.04 $ & $   -2.64 \pm    0.04 $ & $  -2.04$ & $   0.94$  \\ 
NGC 6171   & $ 248.13$ & $ -13.06$ & $   3.37$ & $  23.01$ & $    6.03 \pm    0.31^e $ & $  -34.68 \pm    0.19^e $ & $   -1.94 \pm    0.04 $ & $   -5.95 \pm    0.04 $ & $  -1.00$ & $   0.56$  \\ 
NGC 6235   & $ 253.36$ & $ -22.17$ & $ 358.92$ & $  13.52$ & $   13.52 \pm    1.35^e $ & $  126.68 \pm    0.33^e $ & $   -3.94 \pm    0.04 $ & $   -7.56 \pm    0.04 $ & $  -1.37$ & $   0.33$  \\ 
NGC 6256$^\star$    & $ 254.88$ & $ -37.12$ & $ 347.79$ & $   3.31$ & $    6.40 \pm    0.64^e $ & $ -103.40 \pm    0.50^i $ & $   -3.54 \pm    0.05 $ & $   -1.54 \pm    0.05 $ & $  -1.51$ & $   0.02$  \\ 
NGC 6266$^\star$   & $ 255.31$ & $ -30.11$ & $ 353.58$ & $   7.32$ & $    6.41 \pm    0.12^e $ & $  -73.49 \pm    0.70^e $ & $   -5.06 \pm    0.07 $ & $   -2.98 \pm    0.07 $ & $  -1.09$ & $   0.22$  \\ 
NGC 6273   & $ 255.66$ & $ -26.27$ & $ 356.87$ & $   9.38$ & $    8.27 \pm    0.41^e $ & $  145.54 \pm    0.59^e $ & $   -3.22 \pm    0.04 $ & $    1.61 \pm    0.04 $ & $  -1.70$ & $   0.43$  \\ 
NGC 6284   & $ 256.12$ & $ -24.76$ & $ 358.35$ & $   9.94$ & $   15.30 \pm    1.53^d $ & $   20.70 \pm    0.40^i $ & $   -3.19 \pm    0.04 $ & $   -2.05 \pm    0.04 $ & $  -1.07$ & $   0.07$  \\ 
NGC 6287   & $ 256.29$ & $ -22.71$ & $   0.13$ & $  11.02$ & $    9.40 \pm    0.94^d $ & $ -294.74 \pm    1.65^e $ & $   -4.89 \pm    0.04 $ & $   -1.92 \pm    0.04 $ & $  -2.09$ & $   0.29$  \\ 
NGC 6293   & $ 257.55$ & $ -26.58$ & $ 357.62$ & $   7.83$ & $    9.23 \pm    0.70^e $ & $ -143.66 \pm    0.39^e $ & $    0.82 \pm    0.04 $ & $   -4.31 \pm    0.04 $ & $  -1.92$ & $   0.05$  \\ 
NGC 6304$^\star$   & $ 258.63$ & $ -29.46$ & $ 355.83$ & $   5.38$ & $    6.28 \pm    0.11^f $ & $ -108.62 \pm    0.39^e $ & $   -4.01 \pm    0.06 $ & $   -1.03 \pm    0.05 $ & $  -0.45^f$ & $   0.21$  \\ 
NGC 6316$^\star$    & $ 259.16$ & $ -28.14$ & $ 357.18$ & $   5.76$ & $   11.60 \pm    1.16^e $ & $   92.20 \pm    0.60^i $ & $   -4.86 \pm    0.06 $ & $   -4.61 \pm    0.06 $ & $  -0.46$ & $   0.17$  \\ 
NGC 6325$^\star$    & $ 259.50$ & $ -23.77$ & $   0.97$ & $   8.00$ & $    7.80 \pm    0.78^d $ & $   29.54 \pm    0.58^e $ & $   -8.41 \pm    0.04 $ & $   -8.90 \pm    0.05 $ & $  -1.30$ & $   0.03$  \\ 
NGC 6333   & $ 259.80$ & $ -18.52$ & $   5.54$ & $  10.70$ & $    8.40 \pm    0.84^e $ & $  310.75 \pm    2.12^e $ & $   -2.20 \pm    0.04 $ & $   -3.21 \pm    0.04 $ & $  -1.70$ & $   0.45$  \\ 
NGC 6342$^\star$   & $ 260.29$ & $ -19.58$ & $   4.90$ & $   9.73$ & $    8.43 \pm    0.84^e $ & $  116.56 \pm    0.74^e $ & $   -2.94 \pm    0.05 $ & $   -7.08 \pm    0.05 $ & $  -0.60^m$ & $   0.05$  \\ 
NGC 6352   & $ 261.37$ & $ -48.43$ & $ 341.42$ & $  -7.17$ & $    5.89 \pm    0.58^e $ & $ -123.70 \pm    0.30^i $ & $   -2.19 \pm    0.04 $ & $   -4.42 \pm    0.04 $ & $  -0.48^f$ & $   0.83$  \\ 
NGC 6355$^\star$   & $ 260.99$ & $ -26.36$ & $ 359.58$ & $   5.43$ & $    8.70 \pm    0.87^e$ & $ -210.30 \pm    0.40^i $ & $   -4.66 \pm    0.06 $ & $   -0.51 \pm    0.05 $ & $  -1.46$ & $   0.05$  \\ 
NGC 6356   & $ 260.89$ & $ -17.82$ & $   6.72$ & $  10.22$ & $   15.10 \pm    1.51^d $ & $   38.93 \pm    1.88^e $ & $   -3.77 \pm    0.04 $ & $   -3.37 \pm    0.04 $ & $  -0.51$ & $   0.24$  \\ 
NGC 6362$^\star$    & $ 262.97$ & $ -67.05$ & $ 325.55$ & $ -17.57$ & $    7.69 \pm    0.13^f $ & $  -14.58 \pm    0.18^e $ & $   -5.47 \pm    0.05 $ & $   -4.72 \pm    0.05 $ & $  -1.07^f$ & $   1.13$  \\ 
NGC 6366   & $ 261.93$ & $  -5.08$ & $  18.41$ & $  16.04$ & $    3.66 \pm    0.20^e $ & $ -118.30 \pm    0.30^i $ & $   -0.38 \pm    0.04 $ & $   -5.13 \pm    0.04 $ & $  -0.59$ & $   2.17$  \\ 
NGC 6380   & $ 263.62$ & $ -39.07$ & $ 350.18$ & $  -3.42$ & $    9.80 \pm    0.98^e $ & $   -6.54 \pm    1.48^e $ & $   -2.10 \pm    0.04 $ & $   -3.19 \pm    0.04 $ & $  -0.44$ & $   0.34$  \\ 
NGC 6388   & $ 264.08$ & $ -44.73$ & $ 345.56$ & $  -6.74$ & $   10.74 \pm    0.12^e $ & $   82.85 \pm    0.48^e $ & $   -1.35 \pm    0.04 $ & $   -2.71 \pm    0.04 $ & $  -0.51$ & $   0.12$  \\ 
NGC 6401$^\star$    & $ 264.65$ & $ -23.91$ & $   3.45$ & $   3.98$ & $    7.70 \pm    0.77^e $ & $ -115.40 \pm    0.30^i $ & $   -2.79 \pm    0.06 $ & $    1.49 \pm    0.05 $ & $  -1.08$ & $   0.25$  \\ 
NGC 6402   & $ 264.39$ & $  -3.25$ & $  21.32$ & $  14.81$ & $    9.31 \pm    0.44^e $ & $  -60.71 \pm    0.45^e $ & $   -3.61 \pm    0.04 $ & $   -5.04 \pm    0.04 $ & $  -1.28$ & $   0.79$  \\ 
NGC 6440$^\star$    & $ 267.22$ & $ -20.36$ & $   7.73$ & $   3.80$ & $    8.24 \pm    0.82^e $ & $  -54.00 \pm   26.00^k $ & $   -1.11 \pm    0.07 $ & $   -3.80 \pm    0.07 $ & $  -0.50^m$ & $   0.14$  \\ 
NGC 6441   & $ 267.56$ & $ -37.06$ & $ 353.53$ & $  -5.01$ & $   11.83 \pm    0.14^e $ & $   17.27 \pm    0.93^e $ & $   -2.55 \pm    0.04 $ & $   -5.30 \pm    0.04 $ & $  -0.47$ & $   0.13$  \\ 
NGC 6453   & $ 267.71$ & $ -34.60$ & $ 355.72$ & $  -3.87$ & $   11.60 \pm    1.16^d $ & $ -110.30 \pm    0.50^i $ & $    0.07 \pm    0.04 $ & $   -5.85 \pm    0.04 $ & $  -1.55$ & $   0.05$  \\ 
NGC 6496   & $ 269.76$ & $ -44.26$ & $ 348.03$ & $ -10.01$ & $   11.30 \pm    1.13^d $ & $ -134.72 \pm    0.26^e $ & $   -3.03 \pm    0.07 $ & $   -9.20 \pm    0.04 $ & $  -0.60$ & $   0.95$  \\ 
NGC 6517   & $ 270.47$ & $  -8.96$ & $  19.23$ & $   6.76$ & $   10.60 \pm    1.06^d $ & $  -65.10 \pm    0.20^i $ & $   -1.52 \pm    0.04 $ & $   -4.26 \pm    0.04 $ & $  -1.60$ & $   0.06$  \\ 
NGC 6522$^\star$   & $ 270.89$ & $ -30.03$ & $   1.02$ & $  -3.93$ & $    7.40 \pm    0.19^g $ & $  -14.40 \pm    0.50^l $ & $    2.66 \pm    0.06 $ & $   -6.37 \pm    0.06 $ & $  -0.95^l$ & $   0.05$  \\ 
NGC 6528$^\star$    & $ 271.20$ & $ -30.05$ & $   1.14$ & $  -4.17$ & $    7.70 \pm    0.77 $ & $  185.00 \pm   10.00^k $ & $   -2.21 \pm    0.07 $ & $   -5.55 \pm    0.07 $ & $  -0.17^m$ & $   0.13$  \\ 
NGC 6535   & $ 270.96$ & $  -0.29$ & $  27.18$ & $  10.44$ & $    6.50 \pm    0.65^e $ & $ -214.85 \pm    0.46^e $ & $   -4.21 \pm    0.04 $ & $   -2.95 \pm    0.04 $ & $  -1.79$ & $   0.36$  \\ 
NGC 6539$^\star$    & $ 271.21$ & $  -7.58$ & $  20.80$ & $   6.78$ & $    7.85 \pm    0.66^e $ & $   31.30 \pm    0.40^i $ & $   -6.89 \pm    0.05 $ & $   -3.54 \pm    0.05 $ & $  -0.79^m$ & $   0.38$  \\ 
NGC 6540$^\star$    & $ 271.53$ & $ -27.76$ & $   3.29$ & $  -3.31$ & $    5.20 \pm    0.52^e $ & $  -17.98 \pm    0.84^e $ & $   -3.68 \pm    0.05 $ & $   -2.80 \pm    0.06 $ & $  -0.19^m$ & $   0.03$  \\ 
NGC 6541   & $ 272.00$ & $ -43.71$ & $ 349.29$ & $ -11.18$ & $    7.95 \pm    0.37^e $ & $ -163.97 \pm    0.46^e $ & $    0.28 \pm    0.04 $ & $   -8.77 \pm    0.04 $ & $  -1.80$ & $   0.18$  \\ 
NGC 6544   & $ 271.83$ & $ -24.99$ & $   5.84$ & $  -2.20$ & $    3.00 \pm    0.30^d $ & $  -38.12 \pm    0.76^e $ & $   -2.33 \pm    0.04 $ & $  -18.56 \pm    0.04 $ & $  -1.37$ & $   0.05$  \\ 
NGC 6553   & $ 272.32$ & $ -25.90$ & $   5.25$ & $  -3.02$ & $    6.75 \pm    0.22^e $ & $    0.72 \pm    0.40^e $ & $    0.26 \pm    0.04 $ & $   -0.41 \pm    0.04 $ & $  -0.12$ & $   0.53$  \\ 
NGC 6558$^\star$    & $ 272.57$ & $ -31.76$ & $   0.20$ & $  -6.02$ & $    8.26 \pm    0.53^h $ & $ -194.45 \pm    2.10^h $ & $   -1.78 \pm    0.05 $ & $   -4.12 \pm    0.04 $ & $  -1.17^h$ & $   0.03$  \\ 
NGC 6569   & $ 273.41$ & $ -31.83$ & $   0.48$ & $  -6.68$ & $   10.59 \pm    0.80^e $ & $  -49.83 \pm    0.50^e $ & $   -4.13 \pm    0.04 $ & $   -7.26 \pm    0.04 $ & $  -0.73$ & $   0.35$  \\ 
NGC 6624$^\star$    & $ 275.92$ & $ -30.36$ & $   2.79$ & $  -7.91$ & $    8.43 \pm    0.11^f $ & $   54.26 \pm    0.45^e $ & $    0.10 \pm    0.05 $ & $   -6.92 \pm    0.05 $ & $  -0.69^f$ & $   0.06$  \\ 
NGC 6626$^\star$    & $ 276.14$ & $ -24.87$ & $   7.80$ & $  -5.58$ & $    5.34 \pm    0.17^g $ & $   11.11 \pm    0.60^e $ & $   -0.31 \pm    0.06 $ & $   -8.93 \pm    0.06 $ & $  -1.00^m$ & $   0.24$  \\ 
NGC 6637$^\star$    & $ 277.85$ & $ -32.35$ & $   1.72$ & $ -10.27$ & $    8.80 \pm    0.88 $ & $   46.63 \pm    1.45^e $ & $   -5.11 \pm    0.06 $ & $   -5.82 \pm    0.06 $ & $  -0.77^m$ & $   0.33$  \\ 
NGC 6638$^\star$    & $ 277.73$ & $ -25.49$ & $   7.90$ & $  -7.15$ & $   10.32 \pm    1.03^e $ & $    8.63 \pm    2.00^e $ & $   -2.55 \pm    0.05 $ & $   -4.08 \pm    0.05 $ & $  -0.89$ & $   0.22$  \\ 
NGC 6642$^\star$    & $ 277.97$ & $ -23.48$ & $   9.81$ & $  -6.44$ & $    8.10 \pm    0.81^d $ & $  -33.23 \pm    1.13^e $ & $   -0.23 \pm    0.05 $ & $   -3.88 \pm    0.05 $ & $  -1.03$ & $   0.10$  \\ 
NGC 6652$^\star$    & $ 278.94$ & $ -32.99$ & $   1.53$ & $ -11.38$ & $    9.51 \pm    0.12^f $ & $  -99.04 \pm    0.51^e $ & $   -5.47 \pm    0.05 $ & $   -4.20 \pm    0.05 $ & $  -0.85^f$ & $   0.10$  \\ 
NGC 6656   & $ 138.69$ & $ -49.23$ & $   9.89$ & $  -7.55$ & $    3.20 \pm    0.32^d $ & $ -152.00 \pm   25.00^k $ & $    9.80 \pm    0.04 $ & $   -5.56 \pm    0.04 $ & $  -1.92$ & $   1.33$  \\ 
NGC 6681   & $ 283.26$ & $  -8.70$ & $   2.85$ & $ -12.51$ & $    9.31 \pm    0.17^e $ & $  216.62 \pm    0.84^e $ & $    1.39 \pm    0.04 $ & $   -4.72 \pm    0.04 $ & $  -1.55$ & $   0.03$  \\ 
NGC 6712   & $ 283.27$ & $  -8.70$ & $  25.35$ & $  -4.39$ & $    6.95 \pm    0.39^e $ & $ -107.45 \pm    0.29^e $ & $    3.32 \pm    0.04 $ & $   -4.38 \pm    0.04 $ & $  -0.97$ & $   0.76$  \\ 
NGC 6717$^\star$    & $ 283.78$ & $ -22.70$ & $  12.88$ & $ -10.90$ & $    7.14 \pm    0.10^f $ & $   32.45 \pm    1.44^e $ & $   -3.16 \pm    0.05 $ & $   -4.92 \pm    0.05 $ & $  -1.26^f$ & $   0.08$  \\ 
NGC 6723$^\star$    & $ 284.89$ & $ -36.63$ & $   0.07$ & $ -17.30$ & $    8.17 \pm    0.11^f $ & $  -94.18 \pm    0.26^e $ & $    0.99 \pm    0.05 $ & $   -2.45 \pm    0.05 $ & $  -1.10^f$ & $   0.83$  \\ 
NGC 6752   & $ 287.71$ & $ -59.98$ & $ 336.49$ & $ -25.63$ & $    4.25 \pm    0.09^e $ & $  -26.28 \pm    0.16^e $ & $   -3.17 \pm    0.04 $ & $   -4.01 \pm    0.04 $ & $  -1.41$ & $   0.17$  \\ 
NGC 6760   & $ 287.80$ & $   1.03$ & $  36.11$ & $  -3.92$ & $    7.95 \pm    0.50^e $ & $   -0.42 \pm    1.63^e $ & $   -1.11 \pm    0.04 $ & $   -3.59 \pm    0.04 $ & $  -0.41$ & $   0.34$  \\ 
NGC 6809   & $ 295.00$ & $ -30.97$ & $   8.79$ & $ -23.27$ & $    5.30 \pm    0.20^e $ & $  174.40 \pm    0.24^e $ & $   -3.40 \pm    0.04 $ & $   -9.26 \pm    0.04 $ & $  -1.87$ & $   1.80$  \\ 
NGC 6838   & $ 298.44$ & $  18.78$ & $  56.75$ & $  -4.56$ & $    4.00 \pm    0.20^d $ & $  -42.00 \pm   18.00^k $ & $   -3.38 \pm    0.04 $ & $   -2.65 \pm    0.04 $ & $  -0.63$ & $   0.63$  \\ 
Palomar 11 & $ 296.31$ & $  -8.01$ & $  31.80$ & $ -15.57$ & $   14.30 \pm    1.40^e $ & $  -67.64 \pm    0.76^e $ & $   -1.79 \pm    0.05 $ & $   -4.94 \pm    0.05 $ & $  -0.35$ & $   1.19$  \\ 
Palomar 6$^\star$   & $ 265.93$ & $ -26.22$ & $   2.10$ & $   1.78$ & $    5.80 \pm    0.58^e $ & $  177.00 \pm    1.00^i $ & $   -9.11 \pm    0.06 $ & $   -5.27 \pm    0.06 $ & $  -1.00^m$ & $   0.66$  \\ 
Palomar 7  & $ 272.68$ & $  -7.20$ & $  21.83$ & $   5.67$ & $    5.39 \pm    0.51^e $ & $  155.06 \pm    0.69^e $ & $   -2.47 \pm    0.05 $ & $   -4.41 \pm    0.05 $ & $  -0.53$ & $   1.01$  \\ 
Palomar 8  & $ 280.37$ & $ -19.82$ & $  14.10$ & $  -6.78$ & $   12.80 \pm    1.28^e $ & $  -41.14 \pm    1.81^e $ & $   -2.04 \pm    0.04 $ & $   -5.64 \pm    0.04 $ & $  -0.34$ & $   0.56$  \\ 
Terzan 1$^\star$    & $ 263.95$ & $ -30.47$ & $ 357.57$ & $   1.00$ & $    6.70 \pm    0.67 $ & $   63.00 \pm    0.50^i $ & $   -3.02 \pm    0.05 $ & $   -4.49 \pm    0.04 $ & $  -1.26^m$ & $   0.04$  \\ 
Terzan 10  & $ 270.74$ & $ -26.06$ & $   4.49$ & $  -1.99$ & $   10.40 \pm    1.00^a $ & $  208.60 \pm    3.20^e $ & $   -7.02 \pm    0.07 $ & $   -2.51 \pm    0.06 $ & $  -0.97$ & $   0.90$  \\ 
Terzan 12  & $ 273.06$ & $ -22.74$ & $   8.36$ & $  -2.10$ & $    3.40 \pm    0.34^e $ & $   94.77 \pm    0.97^e $ & $   -6.07 \pm    0.07 $ & $   -2.63 \pm    0.07 $ & $  -0.47$ & $   0.83$  \\ 
Terzan 2$^\star$    & $ 261.89$ & $ -30.80$ & $ 356.32$ & $   2.30$ & $    7.50 \pm    0.75^d $ & $  144.60 \pm    0.80^d $ & $   -2.21 \pm    0.06 $ & $   -6.16 \pm    0.05 $ & $  -0.42$ & $   0.03$  \\ 
Terzan 3$^\star$    & $ 247.17$ & $ -35.35$ & $ 345.08$ & $   9.19$ & $    8.20 \pm    0.82^d $ & $ -135.76 \pm    0.57^e $ & $   -5.61 \pm    0.04 $ & $   -1.69 \pm    0.04 $ & $  -0.96$ & $   1.18$  \\ 
Terzan 4   & $ 262.66$ & $ -31.59$ & $ 356.02$ & $   1.31$ & $    6.70 \pm    0.67^e $ & $  -39.93 \pm    3.76^e $ & $   -5.60 \pm    0.07 $ & $   -2.60 \pm    0.12 $ & $  -1.60^m$ & $   0.90$  \\ 
Terzan 5$^\star$    & $ 267.02$ & $ -24.81$ & $   3.81$ & $   1.67$ & $    5.50 \pm    0.51^e $ & $  -81.40 \pm    1.36^e $ & $   -1.35 \pm    0.07 $ & $   -4.44 \pm    0.06 $ & $  -0.40$ & $   0.16$  \\ 
Terzan 6$^\star$    & $ 267.69$ & $ -31.28$ & $ 358.57$ & $  -2.16$ & $    6.80 \pm    0.68^d $ & $  137.15 \pm    1.70^e $ & $   -5.70 \pm    0.11 $ & $   -7.20 \pm    0.11 $ & $  -0.53$ & $   0.05$  \\ 
Terzan 9$^\star$    & $ 270.41$ & $ -26.84$ & $   3.61$ & $  -1.99$ & $    7.10 \pm    0.71^d $ & $   71.40 \pm    0.40^i $ & $   -2.19 \pm    0.06 $ & $   -7.47 \pm    0.05 $ & $  -1.08$ & $   0.03$  \\ 
Ton 2$^\star$       & $ 264.04$ & $ -38.55$ & $ 350.80$ & $  -3.42$ & $    6.40 \pm    0.64^e $ & $ -172.70 \pm    0.30^i $ & $   -5.92 \pm    0.05 $ & $   -0.55 \pm    0.05 $ & $  -0.26$ & $   0.54$  \\ 
 \\ 
\hline
\hline
\\
\multicolumn{10}{p{17.5cm}}{{\footnotesize{ \textbf{References.} $^a$ \citet{Ortolani+19a} ; $^b$ \citet{Ortolani+19b}; $^c$ \citet{Kerber+19}; $^d$ \citet[updated in 2010]{Harris96}, $^e$ \citet{Baumgardt+19}; $^f$ Oliveira et al. (2019, in preparation); $^g$ \citet{Kerber+18}; $^h$ \citet{Barbuy+18b}; $^i$ \citet{Vasquez+18}; $^j$ \citet{Barbuy+16}; $^k$ \citet{Dias+16}; $^l$ \citet{Barbuy+14}; $^m$\citet[their Table 3]{Bica+2016}. $^\dag$Metallicities with no reference come from the website by Bruno Dias (http://www.sc.eso.org/$\sim$bdias/catalogues.html). $^\ddag$ $R_c$ taken from H10.}}}

\end{longtable}
\normalsize

%% file: TDistance_GCs-2.tex
\begin{longtable}{lccccccc}
\caption{Globular Cluster distances from the literature.}
\label{tab:GC_dist}
\\
\hline
   Cluster & Adopted & H10 & Baumgardt+19 & Valenti+07,10 & Gaia (1/plx)  & Our Group  & Bica+06 \\
    & (kpc) & (kpc) & (kpc) & (kpc)& (kpc) & (kpc)& (kpc)  \\ \endfirsthead
\hline
\hline  
BH 261 &	$6.50\pm 0.65$ & 6.50 &	6.50 & -- & -- & -- & -- \\
Djorg 1* & $9.30\pm0.50$ & 13.70 & 13.70 & 13.50 & -- & 9.30 & 8.80 \\
Djorg 2* & $8.75\pm0.20$ & 6.30 & 6.30 & 7.0 & 9.60 & 8.75 & 6.70 \\
ESO452SC11*	&$6.50\pm0.65$&	8.30&	6.50&	--&	--&	--&	7.8\\
HP 1 &$ 6.59\pm0.16$ & 8.20 & 6.80 & 6.80 & 50.0 & 6.59 & 7.40 \\
Liller 1 & $8.20\pm0.82$ & 8.20 & 8.10 & 7.90 & -- & 7.60 & 10.50 \\
Lynga 7&	$8.0\pm0.80$&	8.0&	8.0&	--&	--&	--&	7.2\\
Mercer 5	&$5.50\pm0.55$&	--	&5.50&	--&	--&	--&	--\\
NGC 104 & $4.50\pm0.45$ & 4.50 & 4.43 & -- & 5.10 & -- & 4.50 \\
NGC 5927* & $8.16\pm0.27$ & 7.70 & 8.16 & -- & 10.04 & -- & 7.60 \\
NGC 6139 &	$9.80\pm0.83$&	10.10&	9.8&	--&	--&	--&	10.1\\
NGC 6144 & $8.90\pm0.89$ & 8.90 & 8.90 & -- & 14.97 & -- & 10.30 \\
NGC 6171 & $6.03\pm0.31$ & 6.40 & 6.03 & -- & 6.76 & -- & 6.40 \\  
NGC 6235* & $13.52\pm1.35$ & 11.50 & 13.52 & -- & 16.18 & -- & 10.00 \\
NGC 6256* & $6.40\pm0.64 $& 10.30 & 6.40 & 9.10 & -- & 6.4 & 6.60 \\
NGC 6266 & $6.41\pm0.12$ & 6.80 & 6.41 & 6.6 & 4.57 & -- & 6.90 \\
NGC 6273 & $8.27\pm0.41$ & 8.80 & 8.27 & 8.20 & 10.82 & -- & 8.70 \\
NGC 6284 & $15.3\pm1.53$ & 15.3 & 15.14 & -- & 20.04 & -- & 14.70 \\
NGC 6287 & $9.40\pm0.94$ & 9.40 & 9.40 & -- & 9.31 & -- & 8.50 \\
NGC 6293 & $9.23\pm0.70$ & 9.50 & 9.23 & 10.50 & 14.37 & -- & 8.80 \\
NGC 6304 & $6.28\pm0.11$ & 5.9 & 5.77 & 6.0 & 9.29 & 6.19 & 6.10 \\
NGC 6316* & $11.60\pm1.16$ & 10.40 & 11.60 & 11.6 & 15.17 & -- & 11.00 \\
NGC 6325 & $7.80\pm0.78$ & 7.80 & 7.80 & -- & 6.99 & 6.90 & 9.60 \\
NGC 6333 & $8.40\pm0.84$ & 7.90 & 8.40 & -- & 10.70 & -- & 8.20 \\
NGC 6342 & $8.43\pm0.84$ & 8.50 & 8.43 & 8.40 & 10.28 & -- & 8.60 \\
NGC 6352 & $5.89\pm0.58$ & 5.60 & 5.89 & -- & 6.48 & -- & 5.70 \\
NGC 6355 & $8.70\pm0.87$ & 9.20 & 8.70 & 9.0 & -- & 8.80 & 7.20 \\
NGC 6356 & $15.10\pm1.51$ & 15.10 & 15.10 & -- & 12.64 & 15 & 15.20 \\
NGC 6362 & $7.69\pm0.13$ & 7.6 & 7.36 & -- & 10.27 & 7.71 & 8.10 \\
NGC 6366 & $3.66\pm0.20$ & 3.5 & 3.66 & -- & 4.36 & -- & 3.60 \\
NGC 6380* & $9.80\pm0.98$ & 10.90 & 9.80 & 9.20 & 9.86 & 9.8 & 10.70 \\
NGC 6388* & $10.74\pm0.12$ & 9.90 & 10.74 & 11.9 & 20.74 & -- & 11.50 \\
NGC 6401* & $7.7\pm0.77$ & 10.6 & 7.7 & 7.7 & 8.65 & 12.0 & 7.70 \\
NGC 6402 & $9.31\pm0.44$ & 9.30 & 9.31 & -- & 18.66 & -- & 8.90 \\
NGC 6440 & $8.24\pm0.82$ & 8.5 & 8.24 & 8.20 & 10.43 & 8.47 & 8.40 \\
NGC 6441 & $11.83\pm0.14$ & 11.60 & 11.83 & 13.5 & 24.81 & -- & 11.20 \\
NGC 6453 &$ 11.60\pm1.16 $ & 11.60 & 11.60 & 10.7 & 23.53 & 8.5 & 11.20 \\
NGC 6496 &$ 11.30\pm1.13 $ & 11.30 & 11.30 & -- & 12.45 & -- & 6.60 \\
NGC 6517 &$ 10.60\pm1.06 $ & 10.60 & 10.60 & -- & 46.08 & -- & 10.80 \\
NGC 6522 &$ 7.40 \pm 0.19$& 7.70 & 8 & 7.40 & 14.35 & 7.40 & 7.80 \\
NGC 6528 &$ 7.70 \pm 0.77$& 7.90 & 7.45 & 7.50 & 13.40 & 7.50 & 9.10 \\
NGC 6535 &$ 6.50 \pm 0.65$& 6.80 & 6.50 & -- & 7.73 & -- & 6.70 \\
NGC 6539 &$ 7.85 \pm 0.66$& 7.8 & 7.85 & 8.4 & 15.87 & -- & 8.40 \\
NGC 6540* & $5.20\pm0.52$ & 5.30 & 5.20 & 5.20 & -- & 3.50 & 3.70 \\
NGC 6541  & $7.95\pm0.37$ & 7.5 & 7.95 & -- & 8.78 & -- & 7.00 \\
NGC 6544  & $3.0\pm0.30 $ & 3.0 & 2.60 & 2.80 & 3.02 & -- & 2.60 \\
NGC 6553* & $6.75\pm0.22$ & 6.0 & 6.75 & 4.90 & -- & 5.20 & 5.60 \\
NGC 6558 & $8.26\pm0.53$ & 7.40 & 7.20 & -- & -- & 8.26 & 7.40 \\
NGC 6569&  $10.59\pm0.80$ &	10.90&	10.59&	12.0&	--&	--&	8.7\\
NGC 6624 & $8.43\pm 0.11$ & 7.9 & 7.19 & 8.4 & 7.26 & 8.32 & 8.00 \\
NGC 6626 & $5.34\pm0.17$ & 5.5 & 5.43 & -- & 6.81 & 5.34 & 5.70 \\
NGC 6637 & $8.80\pm0.88$ & 8.80 & 8.80 & 9.40 & 13.40 & 8.85 & 8.60 \\
NGC 6638*& $10.32\pm1.03$ & 9.40 & 10.32 & 13.5 & -- & -- & 8.40 \\
NGC 6642 & $8.10\pm0.81$ & 8.10 & 8.05 & 7.0 & -- & -- & 7.70 \\
NGC 6652 & $9.51 \pm0.12$ & 10 & 10 & 6.8 & 9.09 & 9.44 & 9.60 \\
NGC 6656 & $3.20\pm0.32$ & 3.20 & 3.23 & 7.90 & 3.84 & -- & 3.20 \\
NGC 6681 & $9.31\pm0.17$ &	9.0&	9.31&	--&	9.12&	--&	--\\
NGC 6712&  $6.95\pm0.39$&	6.90&	6.12&	--&	--&	--&	6.9\\
NGC 6717 & $7.14\pm0.10$ & 7.10 & 7.10 & -- & 8.0 & 7.15 & 7.40 \\
NGC 6723 & $8.17\pm0.11$ & 8.70 & 8.30 & -- & 13.33 & 8.05 & 8.80 \\
NGC 6752&	$4.25\pm0.09$&	4.0&	4.25&	--&	4.32&	4.36&4.0\\
NGC 6760&	$7.95\pm0.50$&	7.40&	7.95&	--&	--&	--&	7.40\\
NGC 6809 & $5.30\pm0.20$ & 5.40 & 5.30 & -- & 5.86 & -- & 5.40 \\
NGC 6838 & $4.0\pm0.20$ & 4.0 & 3.99 & -- & 4.44 & -- & 3.90 \\
Palomar 11&	$14.30\pm1.40$&	13.4&	14.30&	--&	--&	--&	12.9\\
Palomar 6* & $5.80\pm0.58$ & 5.80 & 5.80 & -- & -- & 8.90 & 7.30 \\
Palomar 7&	$5.39\pm0.51$	&5.40&	5.38&	--&	--&	--&	5.4\\
Palomar 8&	$12.80\pm1.28$&	12.80&	12.80&	--&	--&	--&	12.9\\
Terzan 1 & $6.70\pm0.67$ & 6.70 & 6.7 & 9.10 & -- & 5.20 & 6.20 \\
Terzan 10 & $10.3\pm1.0$ & 5.8 & 5.8 & 6.6 & -- & 10.30 & 5.70 \\
Terzan 12&	$3.40\pm0.34$&	4.80&	3.40&	-&	-&	-&	4.80\\
Terzan 2 & $7.50\pm0.75$ & 7.50 & 7.50 & 8.20 & -- & 7.70 & 8.70 \\
Terzan 3 & $8.20\pm0.82$  & 8.20 & 8.10 & -- & -- & 6.50 & 7.50 \\
Terzan 4* & $6.70\pm0.67$ & 7.20 & 6.70 & -- & -- & 8.0 & 9.10 \\
Terzan 5* & $5.50\pm0.51$ & 6.90 & 5.50 & 10.50 & -- & 5.50 & 7.60 \\
Terzan 6 & $6.80\pm0.68$ & 6.80 & 6.70 & 6.0 & -- & 7.0 & 9.50 \\
Terzan 9* & $7.10\pm0.71$ & 7.10 & 7.10 & 11.60 & -- & 4.90 & 7.70 \\
Ton 2 & $6.40\pm0.64$ & 8.20 & 6.40 & -- & -- & 6.40 & 8.10 \\
\hline
\hline
\\
\multicolumn{8}{p{16cm}}{{\footnotesize{* GCs with the largest difference in distance determinations in the literature.}}}
\end{longtable}

%% file: TOrbital_para-2.tex
\clearpage
\onecolumn
\begin{landscape}
\small\addtolength{\tabcolsep}{-0.3pt}
\begin{longtable}{|l |c c c c| c c c c |c c c c|}
\caption{Orbital parameters for different bar pattern speed.}
\label{tab:dynP}
\\
\hline
 & \multicolumn{4}{|c|}{$\Omega_b=40$ $\mathrm{km} \,\mathrm{s}^{-1}\, \mathrm{kpc}^{-1}$} & \multicolumn{4}{|c|}{$\Omega_b=45$ $\mathrm{km} \,\mathrm{s}^{-1}\, \mathrm{kpc}^{-1}$}&\multicolumn{4}{|c|}{$\Omega_b=50$ $\mathrm{km} \,\mathrm{s}^{-1}\, \mathrm{kpc}^{-1}$} \\
 \hline
Cluster & $<r_{\rm min}>$ & $<r_{\rm max}>$ & $<z_{\rm max}>$ & $<e>$ & $<r_{\rm min}>$ & $<r_{\rm max}>$ & $<z_{\rm max}>$ & $<e>$ & $<r_{\rm min}>$ & $<r_{\rm max}>$ & $<z_{\rm max}>$ & $<e>$ \\
 &(kpc) & (kpc) & (kpc) & & (kpc) & (kpc) & (kpc) & & (kpc) & (kpc) & (kpc) & \\  \endfirsthead

\multicolumn{13}{c}%
{{\bfseries \tablename\ \thetable{} -- continued}} \\
\hline
Cluster & $<r_{\rm min}>$ & $<r_{\rm max}>$ & $<z_{\rm max}>$ & $<e>$ & $<r_{\rm min}>$ & $<r_{\rm max}>$ & $<z_{\rm max}>$ & $<e>$ & $<r_{\rm min}>$ & $<r_{\rm max}>$ & $<z_{\rm max}>$ & $<e>$ \\
 &(kpc) & (kpc) & (kpc) & & (kpc) & (kpc) & (kpc) & & (kpc) & (kpc) & (kpc) & \\ 
 \hline
 \endhead

\hline
\endfoot

\hline \hline
\endlastfoot

\hline
\hline
   BH 261 &  $0.47 \pm 0.60$ & $3.10 \pm 0.57$ & $ 1.35 \pm 0.17 $ & $ 0.74 \pm 0.20 $ &  $0.42 \pm 0.53$ & $3.16 \pm 0.62$ & $ 1.34 \pm 0.16 $ & $ 0.77 \pm 0.17 $ &  $0.54 \pm 0.52$ & $3.22 \pm 0.69$ & $ 1.25 \pm 0.15 $ & $ 0.73 \pm 0.16 $  \\ 
   Djorg 1 &  $0.52 \pm 0.31$ & $7.02 \pm 2.42$ & $ 3.32 \pm 1.16 $ & $ 0.86 \pm 0.05 $ &  $0.47 \pm 0.31$ & $7.60 \pm 2.59$ & $ 3.15 \pm 1.17 $ & $ 0.87 \pm 0.05 $ &  $0.50 \pm 0.32$ & $7.37 \pm 2.66$ & $ 3.09 \pm 1.24 $ & $ 0.88 \pm 0.05 $  \\ 
   Djorg 2 &  $0.15 \pm 0.12$ & $1.67 \pm 0.04$ & $ 0.55 \pm 0.09 $ & $ 0.83 \pm 0.11 $ &  $0.14 \pm 0.10$ & $1.66 \pm 0.06$ & $ 0.54 \pm 0.16 $ & $ 0.84 \pm 0.10 $ &  $0.09 \pm 0.08$ & $1.67 \pm 0.11$ & $ 0.76 \pm 0.17 $ & $ 0.91 \pm 0.07 $  \\ 
ESO452SC11 &  $0.12 \pm 0.16$ & $3.45 \pm 0.37$ & $ 2.05 \pm 0.12 $ & $ 0.93 \pm 0.06 $ &  $0.12 \pm 0.12$ & $3.46 \pm 0.42$ & $ 2.04 \pm 0.13 $ & $ 0.93 \pm 0.05 $ &  $0.12 \pm 0.14$ & $3.59 \pm 0.47$ & $ 1.99 \pm 0.14 $ & $ 0.93 \pm 0.05 $  \\ 
      HP 1 &  $0.12 \pm 0.08$ & $3.08 \pm 0.37$ & $ 1.83 \pm 0.11 $ & $ 0.93 \pm 0.04 $ &  $0.11 \pm 0.07$ & $3.12 \pm 0.39$ & $ 1.84 \pm 0.12 $ & $ 0.93 \pm 0.04 $ &  $0.11 \pm 0.05$ & $3.00 \pm 0.43$ & $ 1.87 \pm 0.12 $ & $ 0.93 \pm 0.03 $  \\ 
  Liller 1 &  $0.08 \pm 0.18$ & $1.14 \pm 0.41$ & $ 0.09 \pm 0.06 $ & $ 0.88 \pm 0.23 $ &  $0.10 \pm 0.18$ & $1.15 \pm 0.41$ & $ 0.09 \pm 0.06 $ & $ 0.87 \pm 0.23 $ &  $0.12 \pm 0.19$ & $1.13 \pm 0.41$ & $ 0.09 \pm 0.06 $ & $ 0.84 \pm 0.24 $  \\ 
   Lynga 7 &  $1.17 \pm 0.74$ & $5.27 \pm 0.52$ & $ 1.72 \pm 0.25 $ & $ 0.62 \pm 0.15 $ &  $1.73 \pm 0.47$ & $5.54 \pm 0.63$ & $ 1.63 \pm 0.18 $ & $ 0.53 \pm 0.08 $ &  $1.74 \pm 0.37$ & $6.36 \pm 0.53$ & $ 1.45 \pm 0.18 $ & $ 0.58 \pm 0.08 $  \\ 
  Mercer 5 &  $1.83 \pm 0.20$ & $6.48 \pm 1.19$ & $ 0.17 \pm 0.12 $ & $ 0.57 \pm 0.06 $ &  $1.67 \pm 0.21$ & $7.83 \pm 1.04$ & $ 0.16 \pm 0.11 $ & $ 0.65 \pm 0.07 $ &  $1.47 \pm 0.33$ & $7.10 \pm 0.45$ & $ 0.16 \pm 0.15 $ & $ 0.66 \pm 0.06 $  \\ 
   NGC 104 &  $7.22 \pm 0.12$ & $9.43 \pm 0.98$ & $ 4.47 \pm 0.62 $ & $ 0.13 \pm 0.05 $ &  $7.35 \pm 0.20$ & $9.24 \pm 0.61$ & $ 4.35 \pm 0.49 $ & $ 0.12 \pm 0.03 $ &  $7.37 \pm 0.08$ & $9.08 \pm 0.82$ & $ 4.28 \pm 0.56 $ & $ 0.10 \pm 0.04 $  \\ 
  NGC 5927 &  $4.20 \pm 0.21$ & $6.36 \pm 0.23$ & $ 0.83 \pm 0.03 $ & $ 0.21 \pm 0.04 $ &  $3.35 \pm 0.35$ & $6.07 \pm 0.46$ & $ 0.81 \pm 0.05 $ & $ 0.28 \pm 0.07 $ &  $2.98 \pm 0.68$ & $6.11 \pm 0.47$ & $ 0.82 \pm 0.05 $ & $ 0.29 \pm 0.07 $  \\ 
  NGC 6139 &  $0.27 \pm 0.50$ & $4.50 \pm 0.76$ & $ 2.63 \pm 0.31 $ & $ 0.88 \pm 0.14 $ &  $0.21 \pm 0.42$ & $4.67 \pm 0.80$ & $ 2.59 \pm 0.33 $ & $ 0.90 \pm 0.11 $ &  $0.35 \pm 0.46$ & $4.59 \pm 0.95$ & $ 2.55 \pm 0.40 $ & $ 0.86 \pm 0.13 $  \\ 
  NGC 6144 &  $0.98 \pm 0.69$ & $5.69 \pm 1.11$ & $ 4.61 \pm 0.63 $ & $ 0.71 \pm 0.21 $ &  $0.94 \pm 0.55$ & $5.84 \pm 0.88$ & $ 4.86 \pm 0.81 $ & $ 0.73 \pm 0.16 $ &  $0.77 \pm 0.59$ & $5.86 \pm 0.81$ & $ 4.92 \pm 0.79 $ & $ 0.78 \pm 0.16 $  \\ 
  NGC 6171 &  $0.58 \pm 0.25$ & $5.18 \pm 0.42$ & $ 2.70 \pm 0.14 $ & $ 0.81 \pm 0.08 $ &  $0.52 \pm 0.23$ & $5.19 \pm 0.47$ & $ 2.67 \pm 0.13 $ & $ 0.82 \pm 0.07 $ &  $0.52 \pm 0.22$ & $5.04 \pm 0.62$ & $ 2.65 \pm 0.15 $ & $ 0.82 \pm 0.08 $  \\ 
  NGC 6235 &  $4.66 \pm 1.70$ & $19.73 \pm 16.90$ & $ 10.64 \pm 9.66 $ & $ 0.61 \pm 0.08 $ &  $4.68 \pm 1.71$ & $19.28 \pm 16.88$ & $ 10.39 \pm 9.65 $ & $ 0.61 \pm 0.09 $ &  $4.70 \pm 1.69$ & $20.21 \pm 16.90$ & $ 10.83 \pm 9.64 $ & $ 0.60 \pm 0.08 $  \\ 
  NGC 6256 &  $2.00 \pm 0.83$ & $3.23 \pm 0.38$ & $ 0.61 \pm 0.16 $ & $ 0.23 \pm 0.22 $ &  $1.91 \pm 0.78$ & $3.25 \pm 0.45$ & $ 0.61 \pm 0.16 $ & $ 0.27 \pm 0.20 $ &  $1.52 \pm 0.84$ & $3.37 \pm 0.61$ & $ 0.63 \pm 0.16 $ & $ 0.36 \pm 0.20 $  \\ 
  NGC 6266 &  $0.32 \pm 0.14$ & $2.75 \pm 0.16$ & $ 1.22 \pm 0.17 $ & $ 0.80 \pm 0.08 $ &  $0.35 \pm 0.16$ & $2.82 \pm 0.16$ & $ 1.10 \pm 0.14 $ & $ 0.79 \pm 0.09 $ &  $0.38 \pm 0.14$ & $2.93 \pm 0.19$ & $ 1.09 \pm 0.08 $ & $ 0.78 \pm 0.08 $  \\ 
  NGC 6273 &  $0.19 \pm 0.18$ & $6.25 \pm 0.92$ & $ 4.56 \pm 0.60 $ & $ 0.94 \pm 0.06 $ &  $0.20 \pm 0.21$ & $6.17 \pm 0.96$ & $ 4.47 \pm 0.78 $ & $ 0.94 \pm 0.06 $ &  $0.21 \pm 0.20$ & $6.38 \pm 0.97$ & $ 4.82 \pm 0.88 $ & $ 0.94 \pm 0.06 $  \\ 
  NGC 6284 &  $0.26 \pm 0.94$ & $8.84 \pm 1.81$ & $ 7.34 \pm 1.56 $ & $ 0.94 \pm 0.12 $ &  $0.36 \pm 0.94$ & $9.78 \pm 1.72$ & $ 7.70 \pm 1.52 $ & $ 0.92 \pm 0.12 $ &  $0.33 \pm 0.92$ & $9.14 \pm 1.87$ & $ 7.65 \pm 1.73 $ & $ 0.92 \pm 0.12 $  \\ 
  NGC 6287 &  $0.18 \pm 0.19$ & $8.88 \pm 2.06$ & $ 6.99 \pm 2.41 $ & $ 0.96 \pm 0.03 $ &  $0.18 \pm 0.21$ & $9.56 \pm 2.02$ & $ 7.31 \pm 2.43 $ & $ 0.96 \pm 0.03 $ &  $0.19 \pm 0.19$ & $9.51 \pm 2.11$ & $ 7.63 \pm 2.42 $ & $ 0.96 \pm 0.03 $  \\ 
  NGC 6293 &  $0.14 \pm 0.10$ & $4.38 \pm 1.16$ & $ 2.69 \pm 0.64 $ & $ 0.94 \pm 0.04 $ &  $0.15 \pm 0.11$ & $4.35 \pm 1.16$ & $ 2.58 \pm 0.72 $ & $ 0.94 \pm 0.05 $ &  $0.15 \pm 0.10$ & $4.38 \pm 1.25$ & $ 2.52 \pm 0.81 $ & $ 0.94 \pm 0.04 $  \\ 
  NGC 6304 &  $0.96 \pm 0.19$ & $3.13 \pm 0.16$ & $ 0.91 \pm 0.09 $ & $ 0.53 \pm 0.07 $ &  $1.15 \pm 0.19$ & $3.16 \pm 0.17$ & $ 0.91 \pm 0.07 $ & $ 0.47 \pm 0.07 $ &  $1.18 \pm 0.28$ & $3.29 \pm 0.28$ & $ 0.88 \pm 0.08 $ & $ 0.48 \pm 0.12 $  \\ 
  NGC 6316 &  $0.71 \pm 0.86$ & $4.98 \pm 1.76$ & $ 2.35 \pm 0.49 $ & $ 0.76 \pm 0.15 $ &  $0.56 \pm 0.89$ & $4.91 \pm 1.82$ & $ 2.41 \pm 0.49 $ & $ 0.80 \pm 0.15 $ &  $0.59 \pm 0.91$ & $5.01 \pm 1.81$ & $ 2.34 \pm 0.49 $ & $ 0.79 \pm 0.16 $  \\ 
  NGC 6325 &  $0.14 \pm 0.16$ & $3.46 \pm 0.91$ & $ 2.18 \pm 0.32 $ & $ 0.93 \pm 0.08 $ &  $0.14 \pm 0.17$ & $3.53 \pm 0.90$ & $ 2.20 \pm 0.31 $ & $ 0.92 \pm 0.09 $ &  $0.15 \pm 0.18$ & $3.47 \pm 1.09$ & $ 2.23 \pm 0.31 $ & $ 0.92 \pm 0.10 $  \\ 
  NGC 6333 &  $0.50 \pm 0.23$ & $10.73 \pm 1.23$ & $ 6.29 \pm 1.46 $ & $ 0.91 \pm 0.04 $ &  $0.55 \pm 0.25$ & $11.06 \pm 1.29$ & $ 6.54 \pm 1.41 $ & $ 0.91 \pm 0.04 $ &  $0.55 \pm 0.24$ & $11.44 \pm 1.68$ & $ 6.57 \pm 1.41 $ & $ 0.91 \pm 0.04 $  \\ 
  NGC 6342 &  $0.14 \pm 0.33$ & $3.14 \pm 0.68$ & $ 1.77 \pm 0.22 $ & $ 0.91 \pm 0.11 $ &  $0.18 \pm 0.34$ & $3.28 \pm 0.75$ & $ 1.72 \pm 0.23 $ & $ 0.89 \pm 0.12 $ &  $0.20 \pm 0.37$ & $3.32 \pm 0.84$ & $ 1.71 \pm 0.25 $ & $ 0.88 \pm 0.12 $  \\ 
  NGC 6352 &  $3.14 \pm 0.30$ & $4.43 \pm 0.66$ & $ 0.85 \pm 0.06 $ & $ 0.19 \pm 0.05 $ &  $3.12 \pm 0.16$ & $4.84 \pm 0.84$ & $ 0.88 \pm 0.05 $ & $ 0.22 \pm 0.06 $ &  $3.02 \pm 0.14$ & $5.22 \pm 0.19$ & $ 0.90 \pm 0.11 $ & $ 0.27 \pm 0.03 $  \\ 
  NGC 6355 &  $0.14 \pm 0.14$ & $4.03 \pm 1.19$ & $ 2.48 \pm 1.05 $ & $ 0.94 \pm 0.05 $ &  $0.13 \pm 0.14$ & $4.16 \pm 1.20$ & $ 2.51 \pm 1.04 $ & $ 0.94 \pm 0.05 $ &  $0.14 \pm 0.16$ & $4.32 \pm 1.28$ & $ 2.49 \pm 1.15 $ & $ 0.94 \pm 0.06 $  \\ 
  NGC 6356 &  $2.46 \pm 2.18$ & $9.06 \pm 1.78$ & $ 4.49 \pm 1.13 $ & $ 0.58 \pm 0.21 $ &  $2.68 \pm 2.21$ & $8.66 \pm 1.89$ & $ 4.60 \pm 1.16 $ & $ 0.52 \pm 0.22 $ &  $2.62 \pm 2.22$ & $8.83 \pm 2.05$ & $ 4.57 \pm 1.16 $ & $ 0.54 \pm 0.22 $  \\ 
  NGC 6362 &  $2.94 \pm 0.08$ & $6.64 \pm 0.11$ & $ 3.69 \pm 0.05 $ & $ 0.39 \pm 0.02 $ &  $2.81 \pm 0.03$ & $5.44 \pm 0.03$ & $ 3.34 \pm 0.07 $ & $ 0.32 \pm 0.01 $ &  $1.95 \pm 0.05$ & $5.41 \pm 0.05$ & $ 3.33 \pm 0.08 $ & $ 0.47 \pm 0.01 $  \\ 
  NGC 6366 &  $1.85 \pm 0.29$ & $6.49 \pm 0.25$ & $ 2.40 \pm 0.16 $ & $ 0.56 \pm 0.05 $ &  $1.78 \pm 0.17$ & $7.26 \pm 0.30$ & $ 2.53 \pm 0.29 $ & $ 0.61 \pm 0.03 $ &  $1.44 \pm 0.10$ & $6.69 \pm 0.10$ & $ 2.52 \pm 0.17 $ & $ 0.64 \pm 0.02 $  \\ 
  NGC 6380 &  $0.13 \pm 0.06$ & $2.70 \pm 0.92$ & $ 1.64 \pm 0.72 $ & $ 0.92 \pm 0.06 $ &  $0.10 \pm 0.05$ & $2.73 \pm 1.01$ & $ 1.65 \pm 0.65 $ & $ 0.93 \pm 0.04 $ &  $0.09 \pm 0.05$ & $2.89 \pm 1.05$ & $ 1.61 \pm 0.64 $ & $ 0.94 \pm 0.03 $  \\ 
  NGC 6388 &  $0.17 \pm 0.09$ & $5.96 \pm 0.59$ & $ 3.13 \pm 0.65 $ & $ 0.94 \pm 0.03 $ &  $0.64 \pm 0.20$ & $7.08 \pm 0.42$ & $ 3.24 \pm 0.40 $ & $ 0.84 \pm 0.05 $ &  $0.88 \pm 0.10$ & $4.77 \pm 0.31$ & $ 2.05 \pm 0.13 $ & $ 0.69 \pm 0.04 $  \\ 
  NGC 6401 &  $0.13 \pm 0.24$ & $3.11 \pm 0.76$ & $ 1.64 \pm 0.59 $ & $ 0.92 \pm 0.09 $ &  $0.14 \pm 0.24$ & $3.16 \pm 0.79$ & $ 1.58 \pm 0.57 $ & $ 0.92 \pm 0.09 $ &  $0.15 \pm 0.26$ & $3.15 \pm 0.92$ & $ 1.57 \pm 0.58 $ & $ 0.91 \pm 0.10 $  \\ 
  NGC 6402 &  $0.16 \pm 0.16$ & $5.01 \pm 0.49$ & $ 3.07 \pm 0.19 $ & $ 0.94 \pm 0.05 $ &  $0.20 \pm 0.19$ & $5.05 \pm 0.55$ & $ 3.02 \pm 0.19 $ & $ 0.92 \pm 0.06 $ &  $0.23 \pm 0.18$ & $5.14 \pm 1.12$ & $ 2.98 \pm 0.20 $ & $ 0.92 \pm 0.05 $  \\ 
  NGC 6440 &  $0.09 \pm 0.12$ & $1.52 \pm 0.41$ & $ 0.66 \pm 0.34 $ & $ 0.90 \pm 0.11 $ &  $0.09 \pm 0.17$ & $1.51 \pm 0.41$ & $ 0.70 \pm 0.33 $ & $ 0.90 \pm 0.15 $ &  $0.14 \pm 0.15$ & $1.52 \pm 0.46$ & $ 0.69 \pm 0.37 $ & $ 0.84 \pm 0.14 $  \\ 
  NGC 6441 &  $0.54 \pm 0.23$ & $4.38 \pm 0.25$ & $ 1.92 \pm 0.36 $ & $ 0.78 \pm 0.09 $ &  $0.73 \pm 0.23$ & $4.56 \pm 0.25$ & $ 1.86 \pm 0.27 $ & $ 0.73 \pm 0.08 $ &  $0.45 \pm 0.23$ & $4.40 \pm 0.44$ & $ 1.98 \pm 0.14 $ & $ 0.82 \pm 0.08 $  \\ 
  NGC 6453 &  $0.24 \pm 0.75$ & $5.31 \pm 1.79$ & $ 3.56 \pm 1.43 $ & $ 0.90 \pm 0.13 $ &  $0.24 \pm 0.78$ & $5.33 \pm 2.04$ & $ 3.51 \pm 1.43 $ & $ 0.90 \pm 0.13 $ &  $0.31 \pm 0.82$ & $6.32 \pm 2.12$ & $ 3.69 \pm 1.43 $ & $ 0.89 \pm 0.14 $  \\ 
  NGC 6496 &  $3.75 \pm 1.32$ & $13.99 \pm 10.64$ & $ 5.91 \pm 5.49 $ & $ 0.60 \pm 0.09 $ &  $3.91 \pm 1.31$ & $13.16 \pm 10.63$ & $ 5.68 \pm 5.46 $ & $ 0.61 \pm 0.10 $ &  $3.90 \pm 1.27$ & $12.95 \pm 10.61$ & $ 5.67 \pm 5.48 $ & $ 0.59 \pm 0.10 $  \\ 
  NGC 6517 &  $0.14 \pm 0.11$ & $4.83 \pm 0.70$ & $ 2.96 \pm 0.40 $ & $ 0.94 \pm 0.03 $ &  $0.12 \pm 0.13$ & $4.87 \pm 0.94$ & $ 2.87 \pm 0.40 $ & $ 0.95 \pm 0.03 $ &  $0.12 \pm 0.11$ & $4.86 \pm 1.15$ & $ 2.79 \pm 0.44 $ & $ 0.95 \pm 0.03 $  \\ 
  NGC 6522 &  $0.23 \pm 0.11$ & $1.42 \pm 0.10$ & $ 0.96 \pm 0.02 $ & $ 0.72 \pm 0.11 $ &  $0.27 \pm 0.13$ & $1.41 \pm 0.09$ & $ 0.95 \pm 0.02 $ & $ 0.68 \pm 0.12 $ &  $0.31 \pm 0.15$ & $1.40 \pm 0.09$ & $ 0.95 \pm 0.02 $ & $ 0.64 \pm 0.14 $  \\ 
  NGC 6528 &  $0.07 \pm 0.04$ & $1.77 \pm 0.49$ & $ 0.79 \pm 0.27 $ & $ 0.92 \pm 0.03 $ &  $0.08 \pm 0.06$ & $1.77 \pm 0.51$ & $ 0.78 \pm 0.24 $ & $ 0.91 \pm 0.04 $ &  $0.07 \pm 0.06$ & $1.78 \pm 0.51$ & $ 0.78 \pm 0.23 $ & $ 0.92 \pm 0.05 $  \\ 
  NGC 6535 &  $0.26 \pm 0.40$ & $7.77 \pm 1.20$ & $ 4.06 \pm 0.89 $ & $ 0.94 \pm 0.13 $ &  $0.83 \pm 0.44$ & $5.41 \pm 0.82$ & $ 2.70 \pm 0.73 $ & $ 0.73 \pm 0.15 $ &  $0.88 \pm 0.40$ & $5.26 \pm 0.68$ & $ 2.69 \pm 0.81 $ & $ 0.71 \pm 0.14 $  \\ 
  NGC 6539 &  $1.56 \pm 0.46$ & $3.69 \pm 0.58$ & $ 2.42 \pm 0.38 $ & $ 0.38 \pm 0.14 $ &  $1.43 \pm 0.61$ & $4.40 \pm 0.73$ & $ 2.46 \pm 0.42 $ & $ 0.48 \pm 0.19 $ &  $1.39 \pm 0.58$ & $4.81 \pm 0.69$ & $ 2.45 \pm 0.52 $ & $ 0.54 \pm 0.16 $  \\ 
  NGC 6540 &  $0.78 \pm 0.52$ & $3.24 \pm 0.50$ & $ 0.54 \pm 0.16 $ & $ 0.61 \pm 0.13 $ &  $0.75 \pm 0.44$ & $3.25 \pm 0.52$ & $ 0.60 \pm 0.17 $ & $ 0.62 \pm 0.10 $ &  $0.65 \pm 0.40$ & $3.26 \pm 0.56$ & $ 0.59 \pm 0.15 $ & $ 0.65 \pm 0.10 $  \\ 
  NGC 6541 &  $0.90 \pm 0.36$ & $6.00 \pm 0.47$ & $ 3.07 \pm 0.25 $ & $ 0.75 \pm 0.10 $ &  $0.96 \pm 0.35$ & $6.37 \pm 0.90$ & $ 3.13 \pm 0.37 $ & $ 0.74 \pm 0.08 $ &  $0.97 \pm 0.25$ & $7.05 \pm 0.60$ & $ 3.24 \pm 0.21 $ & $ 0.76 \pm 0.07 $  \\ 
  NGC 6544 &  $0.17 \pm 0.14$ & $6.36 \pm 0.53$ & $ 3.72 \pm 0.86 $ & $ 0.95 \pm 0.04 $ &  $0.19 \pm 0.14$ & $7.12 \pm 0.76$ & $ 3.75 \pm 0.83 $ & $ 0.95 \pm 0.04 $ &  $0.17 \pm 0.14$ & $6.37 \pm 0.57$ & $ 3.38 \pm 0.65 $ & $ 0.95 \pm 0.04 $  \\ 
  NGC 6553 &  $0.20 \pm 0.09$ & $2.40 \pm 0.17$ & $ 0.46 \pm 0.03 $ & $ 0.85 \pm 0.05 $ &  $0.29 \pm 0.09$ & $2.36 \pm 0.17$ & $ 0.45 \pm 0.04 $ & $ 0.78 \pm 0.05 $ &  $0.26 \pm 0.14$ & $2.29 \pm 0.16$ & $ 0.44 \pm 0.02 $ & $ 0.79 \pm 0.07 $  \\ 
  NGC 6558 &  $0.10 \pm 0.07$ & $2.20 \pm 0.31$ & $ 1.27 \pm 0.23 $ & $ 0.92 \pm 0.05 $ &  $0.10 \pm 0.06$ & $2.19 \pm 0.30$ & $ 1.29 \pm 0.23 $ & $ 0.91 \pm 0.05 $ &  $0.09 \pm 0.07$ & $2.17 \pm 0.33$ & $ 1.32 \pm 0.23 $ & $ 0.92 \pm 0.05 $  \\ 
  NGC 6569 &  $0.42 \pm 1.01$ & $3.58 \pm 1.12$ & $ 1.52 \pm 0.21 $ & $ 0.79 \pm 0.25 $ &  $0.63 \pm 0.91$ & $3.78 \pm 1.22$ & $ 1.48 \pm 0.20 $ & $ 0.72 \pm 0.21 $ &  $0.59 \pm 0.87$ & $3.71 \pm 1.36$ & $ 1.42 \pm 0.19 $ & $ 0.73 \pm 0.18 $  \\ 
  NGC 6624 &  $0.08 \pm 0.03$ & $1.80 \pm 0.26$ & $ 1.47 \pm 0.04 $ & $ 0.92 \pm 0.03 $ &  $0.08 \pm 0.03$ & $1.84 \pm 0.18$ & $ 1.46 \pm 0.03 $ & $ 0.92 \pm 0.03 $ &  $0.08 \pm 0.03$ & $1.77 \pm 0.11$ & $ 1.46 \pm 0.04 $ & $ 0.92 \pm 0.03 $  \\ 
  NGC 6626 &  $0.14 \pm 0.09$ & $3.30 \pm 0.21$ & $ 1.78 \pm 0.22 $ & $ 0.92 \pm 0.05 $ &  $0.14 \pm 0.13$ & $3.32 \pm 0.27$ & $ 1.72 \pm 0.19 $ & $ 0.92 \pm 0.06 $ &  $0.13 \pm 0.10$ & $3.36 \pm 0.33$ & $ 1.69 \pm 0.25 $ & $ 0.92 \pm 0.05 $  \\ 
  NGC 6637 &  $0.12 \pm 0.39$ & $2.85 \pm 0.72$ & $ 1.82 \pm 0.25 $ & $ 0.92 \pm 0.14 $ &  $0.11 \pm 0.44$ & $2.82 \pm 0.73$ & $ 1.82 \pm 0.24 $ & $ 0.92 \pm 0.17 $ &  $0.11 \pm 0.44$ & $2.87 \pm 0.81$ & $ 1.79 \pm 0.25 $ & $ 0.92 \pm 0.16 $  \\ 
  NGC 6638 &  $0.12 \pm 0.09$ & $3.51 \pm 0.97$ & $ 2.17 \pm 0.48 $ & $ 0.93 \pm 0.03 $ &  $0.11 \pm 0.10$ & $3.59 \pm 1.05$ & $ 2.16 \pm 0.48 $ & $ 0.94 \pm 0.03 $ &  $0.11 \pm 0.10$ & $3.78 \pm 1.09$ & $ 2.13 \pm 0.46 $ & $ 0.94 \pm 0.03 $  \\ 
  NGC 6642 &  $0.10 \pm 0.09$ & $2.35 \pm 0.48$ & $ 1.41 \pm 0.38 $ & $ 0.92 \pm 0.06 $ &  $0.10 \pm 0.11$ & $2.24 \pm 0.52$ & $ 1.38 \pm 0.41 $ & $ 0.92 \pm 0.06 $ &  $0.09 \pm 0.10$ & $2.30 \pm 0.58$ & $ 1.43 \pm 0.42 $ & $ 0.93 \pm 0.06 $  \\ 
  NGC 6652 &  $0.13 \pm 0.08$ & $4.66 \pm 0.54$ & $ 2.89 \pm 0.13 $ & $ 0.94 \pm 0.03 $ &  $0.13 \pm 0.10$ & $4.72 \pm 0.56$ & $ 2.91 \pm 0.15 $ & $ 0.94 \pm 0.04 $ &  $0.16 \pm 0.11$ & $4.82 \pm 0.88$ & $ 2.96 \pm 0.17 $ & $ 0.94 \pm 0.04 $  \\ 
  NGC 6656 &  $4.35 \pm 0.28$ & $38.45 \pm 4.75$ & $ 10.73 \pm 5.56 $ & $ 0.80 \pm 0.03 $ &  $4.33 \pm 0.29$ & $38.37 \pm 4.76$ & $ 11.06 \pm 5.48 $ & $ 0.80 \pm 0.03 $ &  $4.32 \pm 0.28$ & $38.59 \pm 4.75$ & $ 10.72 \pm 5.52 $ & $ 0.80 \pm 0.03 $  \\ 
  NGC 6681 &  $0.16 \pm 0.07$ & $7.85 \pm 0.62$ & $ 6.21 \pm 0.44 $ & $ 0.96 \pm 0.02 $ &  $0.15 \pm 0.07$ & $7.99 \pm 0.99$ & $ 6.43 \pm 0.48 $ & $ 0.96 \pm 0.02 $ &  $0.16 \pm 0.09$ & $8.56 \pm 1.03$ & $ 6.41 \pm 0.75 $ & $ 0.96 \pm 0.02 $  \\ 
  NGC 6712 &  $0.37 \pm 0.19$ & $6.45 \pm 0.86$ & $ 3.89 \pm 0.36 $ & $ 0.90 \pm 0.05 $ &  $0.53 \pm 0.25$ & $7.42 \pm 0.66$ & $ 3.61 \pm 0.30 $ & $ 0.87 \pm 0.06 $ &  $0.54 \pm 0.27$ & $5.94 \pm 0.34$ & $ 2.90 \pm 0.50 $ & $ 0.83 \pm 0.08 $  \\ 
  NGC 6717 &  $0.13 \pm 0.11$ & $3.02 \pm 0.21$ & $ 1.67 \pm 0.14 $ & $ 0.92 \pm 0.06 $ &  $0.27 \pm 0.18$ & $3.32 \pm 0.23$ & $ 1.45 \pm 0.13 $ & $ 0.85 \pm 0.08 $ &  $0.18 \pm 0.15$ & $3.16 \pm 0.23$ & $ 1.46 \pm 0.11 $ & $ 0.89 \pm 0.07 $  \\ 
  NGC 6723 &  $0.52 \pm 0.32$ & $5.29 \pm 0.61$ & $ 3.89 \pm 0.14 $ & $ 0.84 \pm 0.10 $ &  $0.57 \pm 0.35$ & $4.82 \pm 0.93$ & $ 3.92 \pm 0.16 $ & $ 0.81 \pm 0.13 $ &  $0.23 \pm 0.30$ & $4.73 \pm 0.95$ & $ 3.90 \pm 0.18 $ & $ 0.91 \pm 0.11 $  \\ 
  NGC 6752 &  $3.25 \pm 0.11$ & $6.93 \pm 0.12$ & $ 2.46 \pm 0.07 $ & $ 0.36 \pm 0.02 $ &  $2.55 \pm 0.14$ & $6.51 \pm 0.15$ & $ 2.34 \pm 0.05 $ & $ 0.44 \pm 0.03 $ &  $3.52 \pm 0.20$ & $6.38 \pm 0.09$ & $ 2.33 \pm 0.05 $ & $ 0.29 \pm 0.03 $  \\ 
  NGC 6760 &  $1.62 \pm 0.25$ & $6.04 \pm 0.22$ & $ 1.43 \pm 0.57 $ & $ 0.58 \pm 0.06 $ &  $1.51 \pm 0.19$ & $6.11 \pm 0.53$ & $ 1.05 \pm 0.46 $ & $ 0.60 \pm 0.05 $ &  $1.14 \pm 0.22$ & $7.28 \pm 0.32$ & $ 1.10 \pm 0.38 $ & $ 0.73 \pm 0.05 $  \\ 
  NGC 6809 &  $0.92 \pm 0.24$ & $8.19 \pm 0.79$ & $ 5.20 \pm 0.30 $ & $ 0.80 \pm 0.05 $ &  $0.72 \pm 0.38$ & $7.62 \pm 1.02$ & $ 4.94 \pm 0.29 $ & $ 0.84 \pm 0.07 $ &  $1.05 \pm 0.21$ & $8.74 \pm 1.12$ & $ 4.97 \pm 0.35 $ & $ 0.78 \pm 0.05 $  \\ 
  NGC 6838 &  $4.75 \pm 0.86$ & $7.73 \pm 0.14$ & $ 0.73 \pm 0.08 $ & $ 0.24 \pm 0.10 $ &  $4.75 \pm 0.69$ & $7.60 \pm 0.14$ & $ 0.72 \pm 0.06 $ & $ 0.23 \pm 0.07 $ &  $4.73 \pm 0.67$ & $7.68 \pm 0.23$ & $ 0.73 \pm 0.06 $ & $ 0.24 \pm 0.06 $  \\ 
Palomar 11 &  $5.57 \pm 2.53$ & $8.98 \pm 1.59$ & $ 3.90 \pm 0.62 $ & $ 0.23 \pm 0.16 $ &  $5.67 \pm 2.55$ & $8.99 \pm 1.56$ & $ 3.92 \pm 0.62 $ & $ 0.25 \pm 0.17 $ &  $5.62 \pm 2.49$ & $9.60 \pm 1.53$ & $ 4.12 \pm 0.60 $ & $ 0.26 \pm 0.17 $  \\ 
 Palomar 6 &  $0.17 \pm 0.20$ & $4.76 \pm 1.06$ & $ 2.74 \pm 0.55 $ & $ 0.93 \pm 0.06 $ &  $0.15 \pm 0.21$ & $4.71 \pm 1.22$ & $ 2.71 \pm 0.55 $ & $ 0.93 \pm 0.05 $ &  $0.17 \pm 0.24$ & $4.77 \pm 1.53$ & $ 2.67 \pm 0.56 $ & $ 0.93 \pm 0.06 $  \\ 
 Palomar 7 &  $2.86 \pm 0.33$ & $7.89 \pm 0.90$ & $ 0.94 \pm 0.07 $ & $ 0.47 \pm 0.06 $ &  $2.61 \pm 0.62$ & $7.36 \pm 0.48$ & $ 0.88 \pm 0.05 $ & $ 0.45 \pm 0.07 $ &  $3.46 \pm 0.61$ & $7.10 \pm 0.61$ & $ 0.84 \pm 0.05 $ & $ 0.35 \pm 0.07 $  \\ 
 Palomar 8 &  $2.01 \pm 1.67$ & $6.28 \pm 1.27$ & $ 2.20 \pm 0.21 $ & $ 0.51 \pm 0.22 $ &  $2.03 \pm 1.66$ & $6.55 \pm 1.21$ & $ 2.16 \pm 0.24 $ & $ 0.55 \pm 0.22 $ &  $1.82 \pm 1.67$ & $6.31 \pm 1.15$ & $ 2.22 \pm 0.25 $ & $ 0.55 \pm 0.22 $  \\ 
  Terzan 1 &  $0.05 \pm 0.05$ & $1.73 \pm 0.64$ & $ 0.14 \pm 0.01 $ & $ 0.95 \pm 0.06 $ &  $0.05 \pm 0.08$ & $1.70 \pm 0.65$ & $ 0.14 \pm 0.02 $ & $ 0.94 \pm 0.07 $ &  $0.04 \pm 0.08$ & $1.72 \pm 0.66$ & $ 0.14 \pm 0.02 $ & $ 0.95 \pm 0.07 $  \\ 
 Terzan 10 &  $0.57 \pm 0.71$ & $8.82 \pm 5.04$ & $ 6.04 \pm 4.77 $ & $ 0.87 \pm 0.06 $ &  $0.60 \pm 0.71$ & $9.85 \pm 5.11$ & $ 6.11 \pm 4.89 $ & $ 0.87 \pm 0.06 $ &  $0.59 \pm 0.70$ & $9.43 \pm 5.11$ & $ 6.32 \pm 4.83 $ & $ 0.87 \pm 0.06 $  \\ 
 Terzan 12 &  $2.64 \pm 0.44$ & $6.63 \pm 0.82$ & $ 0.83 \pm 0.09 $ & $ 0.43 \pm 0.04 $ &  $2.19 \pm 0.75$ & $7.42 \pm 0.65$ & $ 0.87 \pm 0.09 $ & $ 0.46 \pm 0.09 $ &  $3.29 \pm 0.78$ & $7.10 \pm 0.42$ & $ 0.85 \pm 0.11 $ & $ 0.37 \pm 0.10 $  \\ 
  Terzan 2 &  $0.05 \pm 0.03$ & $1.37 \pm 0.54$ & $ 0.39 \pm 0.23 $ & $ 0.94 \pm 0.03 $ &  $0.05 \pm 0.03$ & $1.35 \pm 0.56$ & $ 0.38 \pm 0.23 $ & $ 0.93 \pm 0.03 $ &  $0.06 \pm 0.04$ & $1.36 \pm 0.57$ & $ 0.38 \pm 0.12 $ & $ 0.93 \pm 0.06 $  \\ 
  Terzan 3 &  $1.32 \pm 0.62$ & $5.13 \pm 0.61$ & $ 2.33 \pm 0.52 $ & $ 0.62 \pm 0.16 $ &  $1.71 \pm 0.35$ & $5.77 \pm 0.67$ & $ 2.21 \pm 0.59 $ & $ 0.56 \pm 0.09 $ &  $1.77 \pm 0.36$ & $6.18 \pm 0.42$ & $ 2.43 \pm 0.46 $ & $ 0.54 \pm 0.09 $  \\ 
  Terzan 4 &  $0.09 \pm 0.13$ & $1.86 \pm 0.57$ & $ 0.60 \pm 0.15 $ & $ 0.91 \pm 0.09 $ &  $0.11 \pm 0.12$ & $1.87 \pm 0.57$ & $ 0.59 \pm 0.14 $ & $ 0.89 \pm 0.06 $ &  $0.09 \pm 0.15$ & $1.89 \pm 0.60$ & $ 0.59 \pm 0.13 $ & $ 0.90 \pm 0.07 $  \\ 
  Terzan 5 &  $0.47 \pm 0.26$ & $3.06 \pm 0.55$ & $ 0.23 \pm 0.06 $ & $ 0.72 \pm 0.06 $ &  $0.48 \pm 0.21$ & $3.04 \pm 0.57$ & $ 0.23 \pm 0.06 $ & $ 0.71 \pm 0.05 $ &  $0.54 \pm 0.23$ & $3.14 \pm 0.61$ & $ 0.24 \pm 0.11 $ & $ 0.70 \pm 0.06 $  \\ 
  Terzan 6 &  $0.06 \pm 0.03$ & $1.86 \pm 0.72$ & $ 0.38 \pm 0.51 $ & $ 0.94 \pm 0.03 $ &  $0.06 \pm 0.03$ & $1.84 \pm 0.76$ & $ 0.39 \pm 0.46 $ & $ 0.94 \pm 0.03 $ &  $0.05 \pm 0.03$ & $1.83 \pm 0.73$ & $ 0.38 \pm 0.46 $ & $ 0.95 \pm 0.03 $  \\ 
  Terzan 9 &  $0.11 \pm 0.05$ & $1.46 \pm 0.47$ & $ 0.32 \pm 0.06 $ & $ 0.87 \pm 0.07 $ &  $0.10 \pm 0.04$ & $1.46 \pm 0.47$ & $ 0.33 \pm 0.05 $ & $ 0.88 \pm 0.08 $ &  $0.12 \pm 0.06$ & $1.47 \pm 0.46$ & $ 0.33 \pm 0.08 $ & $ 0.84 \pm 0.09 $  \\ 
     Ton 2 &  $1.39 \pm 0.68$ & $5.19 \pm 0.85$ & $ 1.98 \pm 0.18 $ & $ 0.57 \pm 0.14 $ &  $1.75 \pm 0.59$ & $5.74 \pm 0.85$ & $ 1.92 \pm 0.18 $ & $ 0.54 \pm 0.10 $ &  $1.88 \pm 0.48$ & $5.93 \pm 0.59$ & $ 1.91 \pm 0.19 $ & $ 0.54 \pm 0.10 $  \\ 

\hline
\end{longtable}
\end{landscape}

\clearpage
\normalsize
\twocolumn

%% file: TMembership.tex
\small\addtolength{\tabcolsep}{-1.5pt}
\begin{longtable}{|l |c c c c| c c c c |c c c c|}
\caption{Membership probability for different bar pattern speed.}
\label{tab:member_prob}
\\
\hline
 & \multicolumn{4}{c|}{$\Omega_b=40$ $\mathrm{km} \,\mathrm{s}^{-1}\, \mathrm{kpc}^{-1}$} & \multicolumn{4}{c|}{$\Omega_b=45$ $\mathrm{km} \,\mathrm{s}^{-1}\, \mathrm{kpc}^{-1}$}&\multicolumn{4}{c|}{$\Omega_b=50$ $\mathrm{km} \,\mathrm{s}^{-1}\, \mathrm{kpc}^{-1}$} \\
 \hline
Cluster & Bulge & Disk$_{\rm\, thick}$  & Halo$_{\rm\, inner}$  & Halo$_{\rm\, outer}$ & Bulge &  Disk$_{\rm\, thick}$  & Halo$_{\rm\, inner}$  & Halo$_{\rm\, outer}$  & Bulge &  Disk$_{\rm\, thick}$  & Halo$_{\rm\, inner}$  & Halo$_{\rm\, outer}$  \\
 &\multicolumn{4}{c|}{\%} & \multicolumn{4}{c|}{\%}  &\multicolumn{4}{c|}{\%} \\  \endfirsthead

\multicolumn{13}{c}%
{{\bfseries \tablename\ \thetable{} -- continued}} \\
\hline
Cluster &  Bulge & Disk$_{\rm\, thick}$  & Halo$_{\rm\, inner}$  & Halo$_{\rm\, outer}$ & Bulge &  Disk$_{\rm\, thick}$  & Halo$_{\rm\, inner}$  & Halo$_{\rm\, outer}$  & Bulge &  Disk$_{\rm\, thick}$  & Halo$_{\rm\, inner}$  & Halo$_{\rm\, outer}$ \\
 &\multicolumn{4}{c|}{\%} & \multicolumn{4}{c|}{\%}  &\multicolumn{4}{c|}{\%} \\ 
 \hline
 \endhead

\hline
\endfoot

\hline \hline
\endlastfoot

\hline
\hline 
      BH 261  & 94.2 &   5.8 &  0.0 &   0.0 & 93.4 &   6.6 &  0.0 &	  0.0 & 93.4 &   6.6 &  0.0 &   0.0 \\
      Djorg 1 &  0.0 &  93.4 &  6.6 &   0.0 &  0.0 &  88.7 & 11.3 &	  0.0 &  0.0 &  93.2 &  6.8 &   0.0 \\
      Djorg 2 & 99.8 &   0.2 &  0.0 &   0.0 & 99.8 &   0.2 &  0.0 &	  0.0 & 99.7 &   0.3 &  0.0 &   0.0 \\
   ESO452SC11 & 59.2 &  40.8 &  0.0 &   0.0 & 59.6 &  40.4 &  0.0 &	  0.0 & 54.3 &  45.7 &  0.0 &   0.0 \\
         HP 1 & 86.5 &  13.5 &  0.0 &   0.0 & 85.4 &  14.6 &  0.0 &	  0.0 & 87.6 &  12.4 &  0.0 &   0.0 \\
     Liller 1 & 99.8 &   0.2 &  0.0 &   0.0 & 99.8 &   0.2 &  0.0 &	  0.0 & 99.8 &   0.2 &  0.0 &   0.0 \\
      Lynga 7 &  0.7 &  99.3 &  0.0 &   0.0 &  0.3 &  99.7 &  0.0 &	  0.0 &  0.0 & 100.0 &  0.0 &   0.0 \\
     Mercer 5 &  0.0 & 100.0 &  0.0 &   0.0 &  0.0 & 100.0 &  0.0 &	  0.0 &  0.0 & 100.0 &  0.0 &   0.0 \\
      NGC 104 &  0.0 &   1.2 & 98.0 &   0.8 &  0.0 &   2.3 & 97.1 &	  0.6 &  0.0 &   3.5 & 96.0 &   0.5 \\
     NGC 5927 &  0.0 & 100.0 &  0.0 &   0.0 &  0.1 &  99.9 &  0.0 &	  0.0 &  0.1 &  99.9 &  0.0 &   0.0 \\
     NGC 6139 &  0.8 &  99.2 &  0.0 &   0.0 &  0.5 &  99.4 &  0.0 &	  0.0 &  0.8 &  99.1 &  0.0 &   0.0 \\
     NGC 6144 &  0.0 &  87.1 & 12.9 &   0.1 &  0.0 &  74.3 & 25.6 &	  0.1 &  0.0 &  71.2 & 28.7 &   0.1 \\
     NGC 6171 &  0.0 &  99.9 &  0.1 &   0.0 &  0.1 &  99.9 &  0.1 &	  0.0 &  0.1 &  99.9 &  0.0 &   0.0 \\
     NGC 6235 &  0.0 &   0.0 &  0.0 & 100.0 &  0.0 &   0.0 &  0.0 &	100.0 &  0.0 &   0.0 &  0.0 & 100.0 \\
     NGC 6256 & 96.0 &   4.0 &  0.0 &   0.0 & 95.9 &   4.1 &  0.0 &	  0.0 & 94.5 &   5.5 &  0.0 &   0.0 \\
     NGC 6266 & 97.7 &   2.3 &  0.0 &   0.0 & 97.7 &   2.3 &  0.0 &	  0.0 & 97.2 &   2.8 &  0.0 &   0.0 \\
     NGC 6273 &  0.0 &  73.9 & 26.1 &   0.1 &  0.0 &  80.3 & 19.6 &	  0.1 &  0.0 &  55.5 & 44.3 &   0.1 \\
     NGC 6284 &  0.0 &   0.0 & 83.8 &  16.1 &  0.0 &   0.0 & 46.7 &	 53.3 &  0.0 &   0.0 & 65.4 &  34.6 \\
     NGC 6287 &  0.0 &   0.0 & 91.7 &   8.3 &  0.0 &   0.0 & 74.7 &	 25.3 &  0.0 &   0.0 & 58.8 &  41.2 \\
     NGC 6293 &  1.1 &  98.9 &  0.0 &   0.0 &  1.7 &  98.2 &  0.0 &	  0.0 &  2.0 &  98.0 &  0.0 &   0.0 \\
     NGC 6304 & 96.3 &   3.7 &  0.0 &   0.0 & 96.0 &   4.0 &  0.0 &	  0.0 & 94.7 &   5.3 &  0.0 &   0.0 \\
     NGC 6316 &  0.4 &  99.6 &  0.0 &   0.0 &  0.4 &  99.6 &  0.0 &	  0.0 &  0.4 &  99.6 &  0.0 &   0.0 \\
     NGC 6325 & 49.3 &  50.7 &  0.0 &   0.0 & 43.3 &  56.7 &  0.0 &	  0.0 & 44.7 &  55.3 &  0.0 &   0.0 \\
     NGC 6333 &  0.0 &   0.0 & 82.9 &  17.1 &  0.0 &   0.0 & 64.9 &	 35.1 &  0.0 &   0.0 & 47.8 &  52.2 \\
     NGC 6342 & 86.6 &  13.4 &  0.0 &   0.0 & 83.7 &  16.3 &  0.0 &	  0.0 & 82.7 &  17.3 &  0.0 &   0.0 \\
     NGC 6352 & 38.3 &  61.7 &  0.0 &   0.0 & 11.7 &  88.3 &  0.0 &	  0.0 &  2.8 &  97.2 &  0.0 &   0.0 \\
     NGC 6355 &  6.7 &  93.3 &  0.0 &   0.0 &  4.1 &  95.9 &  0.0 &	  0.0 &  2.6 &  97.4 &  0.0 &   0.0 \\
     NGC 6356 &  0.0 &   2.3 & 97.2 &   0.6 &  0.0 &   3.6 & 96.0 &	  0.4 &  0.0 &   2.8 & 96.7 &   0.5 \\
     NGC 6362 &  0.0 &  91.9 &  8.1 &   0.0 &  0.0 &  99.5 &  0.5 &	  0.0 &  0.0 &  99.6 &  0.4 &   0.0 \\
     NGC 6366 &  0.0 &  99.7 &  0.3 &   0.0 &  0.0 &  98.5 &  1.5 &	  0.0 &  0.0 &  99.5 &  0.5 &   0.0 \\
     NGC 6380 & 95.8 &   4.2 &  0.0 &   0.0 & 95.4 &   4.6 &  0.0 &	  0.0 & 94.1 &   5.9 &  0.0 &   0.0 \\
     NGC 6388 &  0.0 &  99.3 &  0.7 &   0.0 &  0.0 &  93.9 &  6.0 &	  0.0 &  2.1 &  97.9 &  0.0 &   0.0 \\
     NGC 6401 & 90.3 &   9.7 &  0.0 &   0.0 & 90.1 &   9.9 &  0.0 &	  0.0 & 90.5 &   9.5 &  0.0 &   0.0 \\
     NGC 6402 &  0.0 &  99.9 &  0.1 &   0.0 &  0.0 &  99.9 &  0.1 &	  0.0 &  0.0 &  99.9 &  0.1 &   0.0 \\
     NGC 6440 & 99.8 &   0.2 &  0.0 &   0.0 & 99.8 &   0.2 &  0.0 &	  0.0 & 99.8 &   0.2 &  0.0 &   0.0 \\
     NGC 6441 & 11.0 &  89.0 &  0.0 &   0.0 &  7.0 &  93.0 &  0.0 &	  0.0 &  8.8 &  91.2 &  0.0 &   0.0 \\
     NGC 6453 &  0.0 &  99.4 &  0.6 &   0.0 &  0.0 &  99.4 &  0.6 &	  0.0 &  0.0 &  95.2 &  4.8 &   0.0 \\
     NGC 6496 &  0.0 &   0.0 &  0.9 &  99.1 &  0.0 &   0.0 &  9.4 &	 90.6 &  0.0 &   0.0 & 14.7 &  85.3 \\
     NGC 6517 &  0.1 &  99.9 &  0.1 &   0.0 &  0.1 &  99.9 &  0.1 &	  0.0 &  0.1 &  99.8 &  0.0 &   0.0 \\
     NGC 6522 & 99.8 &   0.2 &  0.0 &   0.0 & 99.8 &   0.2 &  0.0 &	  0.0 & 99.8 &   0.2 &  0.0 &   0.0 \\
     NGC 6528 & 99.7 &   0.3 &  0.0 &   0.0 & 99.7 &   0.3 &  0.0 &	  0.0 & 99.7 &   0.3 &  0.0 &   0.0 \\
     NGC 6535 &  0.0 &  38.9 & 60.9 &   0.2 &  0.0 &  99.9 &  0.1 &	  0.0 &  0.0 &  99.9 &  0.1 &   0.0 \\
     NGC 6539 & 19.5 &  80.5 &  0.0 &   0.0 &  2.2 &  97.8 &  0.0 &	  0.0 &  0.5 &  99.5 &  0.0 &   0.0 \\
     NGC 6540 & 96.1 &   3.9 &  0.0 &   0.0 & 95.9 &   4.1 &  0.0 &	  0.0 & 95.8 &   4.2 &  0.0 &   0.0 \\
     NGC 6541 &  0.0 &  99.4 &  0.6 &   0.0 &  0.0 &  98.6 &  1.4 &	  0.0 &  0.0 &  94.2 &  5.7 &   0.0 \\
     NGC 6544 &  0.0 &  94.6 &  5.4 &   0.0 &  0.0 &  80.7 & 19.3 &	  0.1 &  0.0 &  97.4 &  2.6 &   0.0 \\
     NGC 6553 & 99.3 &   0.7 &  0.0 &   0.0 & 99.4 &   0.6 &  0.0 &	  0.0 & 99.4 &   0.6 &  0.0 &   0.0 \\
     NGC 6558 & 99.1 &   0.9 &  0.0 &   0.0 & 99.1 &   0.9 &  0.0 &	  0.0 & 99.1 &   0.9 &  0.0 &   0.0 \\
     NGC 6569 & 78.0 &  22.0 &  0.0 &   0.0 & 68.5 &  31.5 &  0.0 &	  0.0 & 74.6 &  25.4 &  0.0 &   0.0 \\
     NGC 6624 & 99.3 &   0.7 &  0.0 &   0.0 & 99.3 &   0.7 &  0.0 &	  0.0 & 99.3 &   0.7 &  0.0 &   0.0 \\
     NGC 6626 & 81.1 &  18.9 &  0.0 &   0.0 & 82.4 &  17.6 &  0.0 &	  0.0 & 81.8 &  18.2 &  0.0 &   0.0 \\
     NGC 6637 & 91.7 &   8.3 &  0.0 &   0.0 & 92.2 &   7.8 &  0.0 &	  0.0 & 91.8 &   8.2 &  0.0 &   0.0 \\
     NGC 6638 & 46.7 &  53.3 &  0.0 &   0.0 & 42.3 &  57.7 &  0.0 &	  0.0 & 32.1 &  67.9 &  0.0 &   0.0 \\
     NGC 6642 & 98.5 &   1.5 &  0.0 &   0.0 & 98.8 &   1.2 &  0.0 &	  0.0 & 98.6 &   1.4 &  0.0 &   0.0 \\
     NGC 6652 &  0.2 &  99.8 &  0.0 &   0.0 &  0.1 &  99.8 &  0.0 &	  0.0 &  0.1 &  99.9 &  0.1 &   0.0 \\
     NGC 6656 &  0.0 &   0.0 &  0.0 & 100.0 &  0.0 &   0.0 &  0.0 &	100.0 &  0.0 &   0.0 &  0.0 & 100.0 \\
     NGC 6681 &  0.0 &   0.4 & 98.3 &   1.3 &  0.0 &   0.2 & 97.9 &	  1.9 &  0.0 &   0.1 & 97.5 &   2.4 \\
     NGC 6712 &  0.0 &  90.8 &  9.2 &   0.0 &  0.0 &  77.7 & 22.2 &	  0.1 &  0.0 &  99.6 &  0.4 &   0.0 \\
     NGC 6717 & 91.4 &   8.6 &  0.0 &   0.0 & 88.9 &  11.1 &  0.0 &	  0.0 & 92.1 &   7.9 &  0.0 &   0.0 \\
     NGC 6723 &  0.0 &  98.6 &  1.4 &   0.0 &  0.0 &  99.4 &  0.6 &	  0.0 &  0.0 &  99.5 &  0.5 &   0.0 \\
     NGC 6752 &  0.0 &  99.3 &  0.7 &   0.0 &  0.0 &  99.8 &  0.2 &	  0.0 &  0.0 &  99.8 &  0.2 &   0.0 \\
     NGC 6760 &  0.0 & 100.0 &  0.0 &   0.0 &  0.0 &  99.9 &  0.0 &	  0.0 &  0.0 & 100.0 &  0.0 &   0.0 \\
     NGC 6809 &  0.0 &   2.1 & 97.4 &   0.5 &  0.0 &   9.6 & 90.1 &	  0.3 &  0.0 &   1.3 & 98.1 &   0.5 \\
     NGC 6838 &  0.0 & 100.0 &  0.0 &   0.0 &  0.0 & 100.0 &  0.0 &	  0.0 &  0.0 & 100.0 &  0.0 &   0.0 \\
   Palomar 11 &  0.0 &   9.6 & 89.9 &   0.5 &  0.0 &   9.1 & 90.4 &	  0.5 &  0.0 &   2.0 & 97.1 &   0.9 \\
    Palomar 6 &  0.2 &  99.8 &  0.0 &   0.0 &  0.3 &  99.7 &  0.0 &	  0.0 &  0.3 &  99.7 &  0.0 &   0.0 \\
    Palomar 7 &  0.0 &  99.9 &  0.1 &   0.0 &  0.0 & 100.0 &  0.0 &	  0.0 &  0.0 & 100.0 &  0.0 &   0.0 \\
    Palomar 8 &  0.0 &  99.9 &  0.1 &   0.0 &  0.0 &  99.8 &  0.2 &	  0.0 &  0.0 &  99.9 &  0.1 &   0.0 \\
     Terzan 1 & 99.7 &   0.3 &  0.0 &   0.0 & 99.7 &   0.3 &  0.0 &	  0.0 & 99.7 &   0.3 &  0.0 &   0.0 \\
    Terzan 10 &  0.0 &   0.1 & 98.2 &   1.7 &  0.0 &   0.0 & 95.3 &	  4.7 &  0.0 &   0.0 & 95.9 &   4.1 \\
    Terzan 12 &  0.0 & 100.0 &  0.0 &   0.0 &  0.0 & 100.0 &  0.0 &	  0.0 &  0.0 & 100.0 &  0.0 &   0.0 \\
     Terzan 2 & 99.8 &   0.2 &  0.0 &   0.0 & 99.8 &   0.2 &  0.0 &	  0.0 & 99.8 &   0.2 &  0.0 &   0.0 \\
     Terzan 3 &  0.2 &  99.8 &  0.0 &   0.0 &  0.0 &  99.9 &  0.0 &	  0.0 &  0.0 &  99.8 &  0.2 &   0.0 \\
     Terzan 4 & 99.7 &   0.3 &  0.0 &   0.0 & 99.7 &   0.3 &  0.0 &	  0.0 & 99.7 &   0.3 &  0.0 &   0.0 \\
     Terzan 5 & 97.4 &   2.6 &  0.0 &   0.0 & 97.5 &   2.5 &  0.0 &	  0.0 & 96.9 &   3.1 &  0.0 &   0.0 \\
     Terzan 6 & 99.7 &   0.3 &  0.0 &   0.0 & 99.7 &   0.3 &  0.0 &	  0.0 & 99.7 &   0.3 &  0.0 &   0.0 \\
     Terzan 9 & 99.8 &   0.2 &  0.0 &   0.0 & 99.8 &   0.2 &  0.0 &	  0.0 & 99.8 &   0.2 &  0.0 &   0.0 \\
        Ton 2 &  0.5 &  99.5 &  0.0 &   0.0 &  0.1 &  99.9 &  0.0 &	  0.0 &  0.0 &  99.9 &  0.0 &   0.0 \\
 \hline   
\end{longtable} 
 
\clearpage 
\normalsize
\twocolumn

%% file: TGCs_classification.tex
\begin{table}
\caption{Classification of GCs based on dynamical properties, assuming their maximum membership probability.}
\label{tab:classification}
\begin{tabular}{l|lll}
\hline
\hline
Component & \multicolumn{3}{c}{Clusters} \\  
\hline
\hline

\multirow{10}{*}{Bulge/Bar} & BH 261    &  NGC 6401 & NGC 6637  \\
                            & Djorg 2  &   NGC 6440 & NGC 6642  \\
                            & ESO452SC11 &  NGC 6522 & NGC 6717 \\
                            & HP 1 &  NGC 6528 & Terzan 1  \\
                            &  Liller 1 &  NGC 6540 & Terzan 2  \\
                            &  NGC 6256 &  NGC 6553 & Terzan 4  \\
                            &  NGC 6266 &  NGC 6558 & Terzan 5  \\
                            &  NGC 6304 &  NGC 6569 & Terzan 6  \\
                            &  NGC 6342 &  NGC 6624 & Terzan 9  \\
                            &  NGC 6380 &  NGC 6626 &           \\
\hline   

\multirow{13}{*}{Thick disk} &  Djorg 1 & NGC 6388 &  NGC 6712 \\
                           &    Lynga 7 & NGC 6362 &  NGC 6723 \\
                           &    Mercer 5 & NGC 6366 &  NGC 6752 \\
                           &    NGC 5927 & NGC 6402 &  NGC 6760 \\
                           &    NGC 6139 & NGC 6441 &  NGC 6838 \\
                           &    NGC 6144 & NGC 6453 & Palomar 6 \\
                           &    NGC 6171 & NGC 6517 & Palomar 7 \\
                           &    NGC 6273 & NGC 6535* & Palomar 8 \\
                           &    NGC 6293 & NGC 6539 & Terzan 12 \\
                           &    NGC 6316 & NGC 6541 &  Terzan 3 \\
                           &    NGC 6325 & NGC 6544 &     Ton 2 \\
                           &    NGC 6352 & NGC 6638 &          \\
                           &    NGC 6355 & NGC 6652 &          \\ 
                            
\hline           
\multirow{3}{*}{Inner Halo} & NGC 104  &  NGC 6356 & Palomar 11 \\
                            & NGC 6284* &  NGC 6681 & Terzan 10 \\
                            & NGC 6287 &  NGC 6809 & NGC 6333* \\
\hline 
\multirow{1}{*}{Outer Halo} & NGC 6235 &  NGC 6496 & NGC 6656 \\
\hline
\hline
\end{tabular}
\\
\\
{\footnotesize{*GCs that its classification changes with $\Omega_b$.}}
\end{table}